\documentclass[a4paper,11pt]{article}
\pdfoutput=1

\usepackage{jcappub}
\usepackage[T1]{fontenc}
\usepackage{diagbox}
\usepackage{aas_macros}
\usepackage{xcolor}
\usepackage{orcidlink}

\title{Neutron Capture Information in Improving IBD Angular Resolution}

\author[a,b,c]{Qishan Liu \orcidlink{0000-0003-1437-6829}}
\author[a]{Kenny C. Y. Ng \orcidlink{0000-0001-8016-2170}}

\affiliation[a]{Department of Physics, The Chinese University of Hong Kong, Shatin, New Territories, Hong Kong}
\affiliation[b]{Institute of High Energy Physics, Chinese Academy of Sciences, Beijing 100049, China}
\affiliation[c]{China Center of Advanced Science and Technology, Beijing 100190, China}

\emailAdd{qisliu@link.cuhk.edu.hk}
\emailAdd{kcyng@cuhk.edu.hk}

\abstract{
One of the most important neutrino interactions is the Inverse Beta Decay (IBD). However, the IBD events typically carry no directional information in water Cherenkov detectors as the positron directions are mostly isotropic at low energies, such as those in supernova studies. As gadolinium is being added to Super-Kamiokande, the improved neutron tagging efficiency not only allows better background rejection, but the neutron capture information could potentially provide additional information that allows better event reconstruction. Due to neutron diffusion in water, event-by-event reconstruction is difficult. 
However, if the final neutron capture position is correlated with the initial neutrino momentum, it may be possible that neutrino directionality could be reconstructed statistically, with or without using the positron information.  
In this work, we use Geant4 to simulate neutron propagation in water. We show that in a wide range of neutrino energies from about 10 MeV to several hundred MeV, neutron capture information could statistically enhance the neutrino directionality, compared to positron-only inference, even with neutron diffusion considered.  However, practical application of this technique depends crucially on detection effects, especially the vertex reconstruction resolutions. Our work therefore motivates developments of better reconstruction algorithms and techniques, as well as detector upgrades.
}

\begin{document}
\maketitle
\flushbottom

\section{Introduction} \label{sec:introduction}

While neutrinos are notoriously difficult to detect, low-energy astrophysical neutrino sources, such as the Sun~\cite{Bahcall:1987jc}, supernova SN1987A~\cite{Kamiokande-II:1987idp,Hirata:1988ad,Bionta:1987qt, IMB:1988suc,Alekseev:1988gp}, as well as terrestrial sources such as reactors~\cite{Qian:2018wid}, radioactive nuclei inside the Earth~\cite{Fiorentini:2007te}, and the atmosphere~\cite{Super-Kamiokande:1998kpq,Kajita:2010zz,Zhou:2023mou}, have provided significant new insights into both astrophysics and fundamental physics.

Ideally, one hopes to recover the directions of the incoming neutrinos through various techniques.  The triangulation method, which utilizes multiple detectors at different locations, could be used to determine the direction of supernova neutrinos~\cite{Muhlbeier:2013gwa,Brdar:2018zds,Hansen:2019giq}. 
In the neutrino electron scatterings, the scattering angles can also be utilized to localize a supernova~\cite{Beacom:1998fj,Laha:2013hva,Adams:2013ana}. However, the small cross section of neutrino electron scattering compared to neutrino nucleus scattering limits its effectiveness.

For electron antineutrinos, the Inverse Beta Decay~(IBD, $\bar{\nu}_{e} + p \rightarrow n + e^{+}$) interaction is the dominant reaction in the tens of MeV range~\cite{Formaggio:2012cpf,Ricciardi:2022pru}.
Due to its large cross section and clearly visible positron final state, IBD plays an important role in various key areas of neutrino science, such as supernova neutrinos~\cite{Mirizzi:2015eza}, geoneutrinos~\cite{Araki:2005qa, Fiorentini:2007te}, reactor neutrinos~\cite{JUNO:2025gmd, Qian:2018wid, JUNO:2015zny}, and dark matter~(DM) searches~\cite{Yuksel:2007ac,Palomares-Ruiz:2007trf,Liu:2023cqs,Cappiello:2019qsw,Arguelles:2019ouk,Tse:2026ppi}, etc.  

The distribution of positron angle $\phi$ for IBD can be approximated by the form~\cite{Vogel:1999zy}:
\begin{equation}
\frac{d N}{d \cos \phi}=\frac{N}{2}(1+a \cos \phi),
\end{equation}
where the coefficient $a$ depends on the neutrino energy $E_\nu$. 
Typically, $a$ is small for MeV neutrinos, thus the angular information is lost in positrons, limiting their effectiveness in pinpointing low-energy neutrino sources~\cite{Vogel:1999zy}. 

For pure liquid scintillator detectors, IBD events are characterized by a prompt signal arising primarily from the positron’s kinetic energy deposition followed by a delayed signal from neutron capture on hydrogen. It has been proposed to utilize the displacement between the positron annihilation site and the neutron capture point from IBD events to statistically infer the direction of a supernova~\cite{Beacom:1998fj,Vogel:1999zy,CHOOZ:1999hgz,Fischer:2015oma,Mukhopadhyay:2020ubs,Li:2020gaz}. While this approach has found application in liquid scintillator detectors, its implementation in water Cherenkov detectors remains limited due to the low neutron tagging efficiency in pure water. 

In this paper, we aim to study the possibilities of using neutron capture information to improve the angular resolution of IBD event reconstruction in water Cherenkov detectors. The GADZOOKS~\cite{Beacom:2003nk} proposal pioneered the idea of doping the water medium with gadolinium (Gd) to enhance neutron tagging efficiency. This innovation could potentially facilitate the neutron capture information for reconstructing the direction of IBD events in water Cherenkov detectors such as Super-Kamiokande~(SK)~\cite{Suzuki:2019jby}.  SK is undergoing an upgrade with the ultimate aim of incorporating 0.1\% Gd into the current detector~\cite{Takeuchi:2022dfj,Super-Kamiokande:2021the, Abe:2024ydm, Koshio:2025fjs}.

To use the neutron capture information for reconstructing the direction of IBD neutrinos, several challenges must be addressed, including neutron diffusion in the water, IBD vertex and neutron capture vertex resolutions, and positron angular and energy resolutions.  
We use Geant4 to simulate neutron diffusion and show that neutrons retain some directional information after diffusion. First, we reconstruct the neutrino direction by considering neutron diffusion alone based on the kinematics of IBD. We then take into account both the neutron diffusion and the detector resolutions to estimate the uncertainty of the reconstructed direction. Finally, we discuss the applicability of this method.

\section{Gadolinium Loading in Water Cherenkov Detectors}
SK is a 50-kton water Cherenkov detector in Japan, with a fiducial volume of about 22.5 kton~\cite{Super-Kamiokande:2002weg}. 
To enhance the neutron tagging efficiency, and thus allow neutron tagging as a background removal technique for IBD events~\cite{Li:2014sea,Nikrant:2017nya,Nairat:2024upg}, Gd was proposed to be doped into the medium~\cite{Beacom:2003nk}. 
In 2020, 13 tons of gadolinium sulfate octahydrate (Gd\textsubscript{2}(SO\textsubscript{4})\textsubscript{3}·8H\textsubscript{2}O) were dissolved into the ultrapure water of SK, achieving a Gd concentration of 0.011\% by weight~\cite{Super-Kamiokande:2021the}. This resulted in approximately 50\% of neutrons being captured by Gd. Subsequently, the concentration of Gd was raised to 0.03\% in 2022, increasing the neutron capture efficiency to 75\%~\cite{Super-Kamiokande:2023xup, Abe:2024ydm}. 

The primary objective of SK is to achieve a Gd concentration of 0.1\%, which would significantly enhance neutron capture efficiency to 90\%~\cite{Nakahata:2015vma}. This enhancement is attributed to the substantial thermal neutron capture cross section of Gd, which is 49700 barns, far surpassing the 0.33 barns for hydrogen. Moreover, when a neutron is captured by free protons, it results in gamma-ray emission of only 2.2~MeV. In contrast, the capture of a neutron by Gd triggers the subsequent emission of approximately 8~MeV gamma-ray cascades. These gamma-ray cascades, in coincidence with a prompt positron signal, can uniquely identify electron antineutrinos that interact via IBD, providing a distinctive signature for background suppression~\cite{Renshaw:2012np}.

With such a high efficiency of neutron tagging for IBD events by Gd, we are interested in exploring the potential benefits to better reconstruct the IBD event directionality using the kinematic information provided by the delayed neutron capture. Throughout this paper, we simulate neutron capture at the target Gd concentration of 0.1\%, which provides the benchmark scenario for evaluating this method. Note that our simulation tracks neutron transport and capture at the physics level, but does not model the detector response that determines the neutron tagging efficiency.

\section{Kinematics of Inverse Beta Decay}
\label{sub:kinematics}
\begin{figure}[htbp]
		\centering
		\includegraphics[width= 0.70\columnwidth]{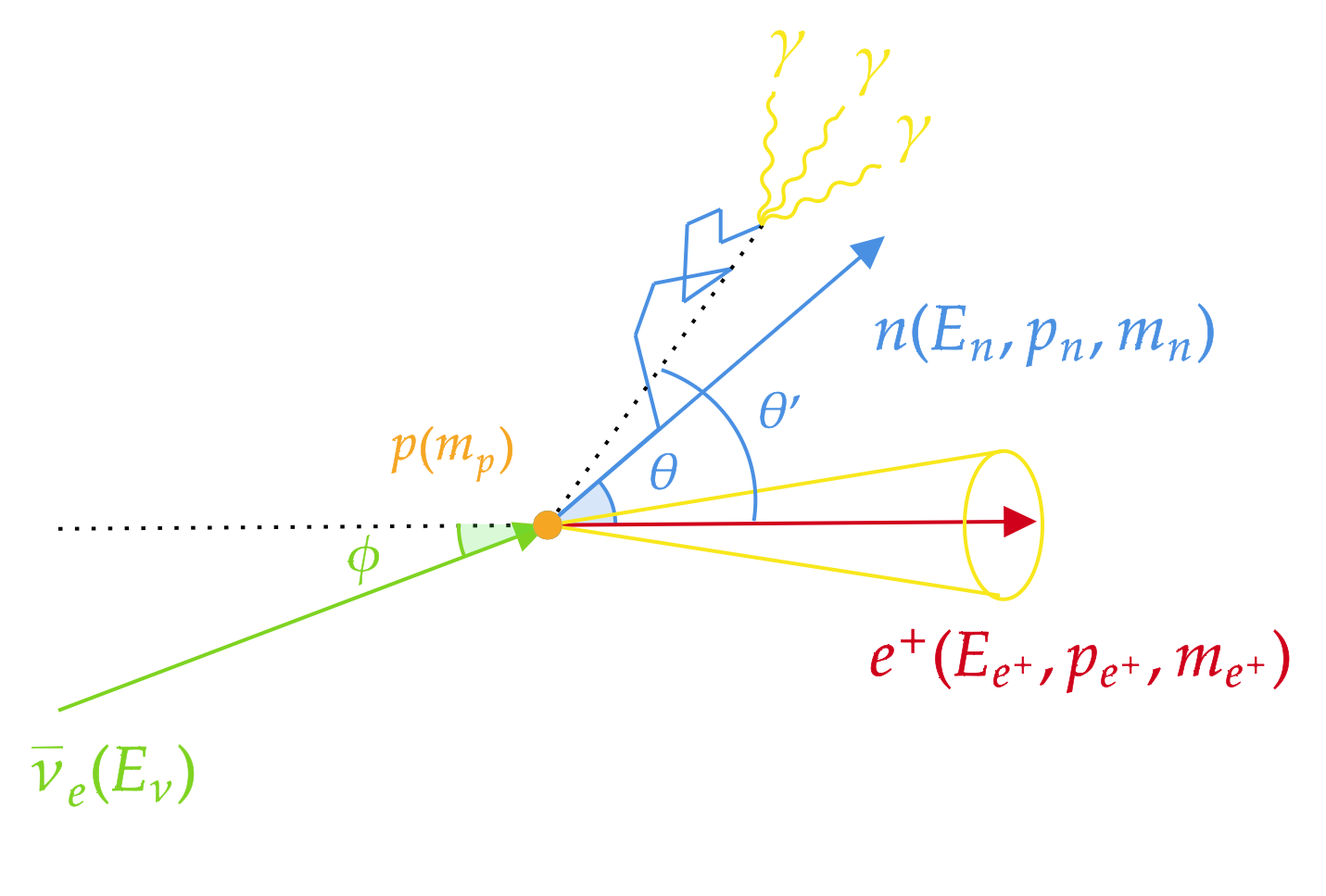}
		\caption{ The IBD process involves an electron antineutrino~(green) interacting with a proton~(orange) at rest. This interaction results in the emission of a positron~(red) and a neutron~(blue). The neutron would diffuse and emit gamma rays~(yellow) after being captured.}
        \label{fig:ibd}
\end{figure}

We first describe the kinematics of IBD, which determines the observables that are needed for the event reconstruction.  
IBD is the charged-current quasielastic reaction of an electron antineutrino scattering with a proton at rest, 
\begin{equation}
\bar{\nu}_e+p \rightarrow e^{+}+n.
\end{equation}
The kinematics of IBD is shown schematically in Fig.~\ref{fig:ibd}. 
We choose our coordinate frame by setting the positron track to be the positive $x$ direction. Then we are free to choose the scattering plane to be the $x-y$ plane.  Ignoring the neutrino mass, the conservation of 4-momentum is then represented as the following sets of equations
\begin{equation}
\left[\begin{array}{c}
E_\nu \\
E_\nu \cos \phi\\
E_\nu \sin \phi \\0
\end{array}\right]+\left[\begin{array}{c}
m_{\mathrm{p}} \\
0 \\0 \\
0
\end{array}\right]=\left[\begin{array}{c}
E_{\mathrm{n}} \\
p_{\mathrm{n}} \cos \theta \\
p_{\mathrm{n}} \sin \theta\\0
\end{array}\right]+\left[\begin{array}{c}
E_e \\
p_e  \\0 \\
0
\end{array}\right]\ , 
\end{equation}
where $E_\nu$, $E_e$, $E_{\mathrm{n}}$ are the energies of neutrino, positron, and neutron, $p_e$ and $p_{\mathrm{n}}$ are the momenta of positron and neutron, $m_{\mathrm{p}}$ is the mass of the proton, $\phi$ is the angle between the neutrino and the positron,  and $\theta$ is the angle between the neutron and the positron.  There are a total of 7 parameters and we have 5 equations, 
\begin{equation}
\label{eq:ibd}
\begin{cases}E_\nu+m_{\mathrm{p}} & =E_{\mathrm{n}}+E_e \\ E_\nu \cos \phi& =p_{\mathrm{n}} \cos \theta+p_e  \\ E_\nu \sin \phi & =p_{\mathrm{n}} \sin \theta \\ m_e^2 & =E_e^2-p_e^2 \\ m_{\mathrm{n}}^2 & =E_{\mathrm{n}}^2-p_{\mathrm{n}}^2\end{cases}\ ,
\end{equation}
with the last two being the on-shell conditions. This means that we need to measure 2 parameters to fully reconstruct the event.  One obvious measurable parameter is the energy of the positron, and the other can be either the momentum of the neutron or the angle $\theta$.  In this work, we focus on the angle $\theta$. 

In practice, neutron diffusion in water~\cite{Super-Kamiokande:2008mmn,Renshaw:2012np} makes it difficult to meaningfully reconstruct at the event-by-event level.  Considering the neutron capture position after diffusion, one would obtain the angle $\theta'$ instead, as shown in Fig.~\ref{fig:ibd}.  However, if some directionality remains, one may \emph{statistically} infer the direction of the neutrinos using multiple events, and this is the theoretically \emph{best} scenario.  Experimental limitations on reconstructing the interaction vertex, the neutron capture position (these two affect the $\theta'$ accuracy) and the energy/angular resolutions of the positron further degrade the final neutrino directionality.  The bulk of this work is to investigate the effects of all these factors. 

\section{Neutron Capture Analysis}

\begin{figure*}[htbp]
  \centering
  
  \begin{minipage}[b]{0.325\textwidth}
    \includegraphics[width=\textwidth]{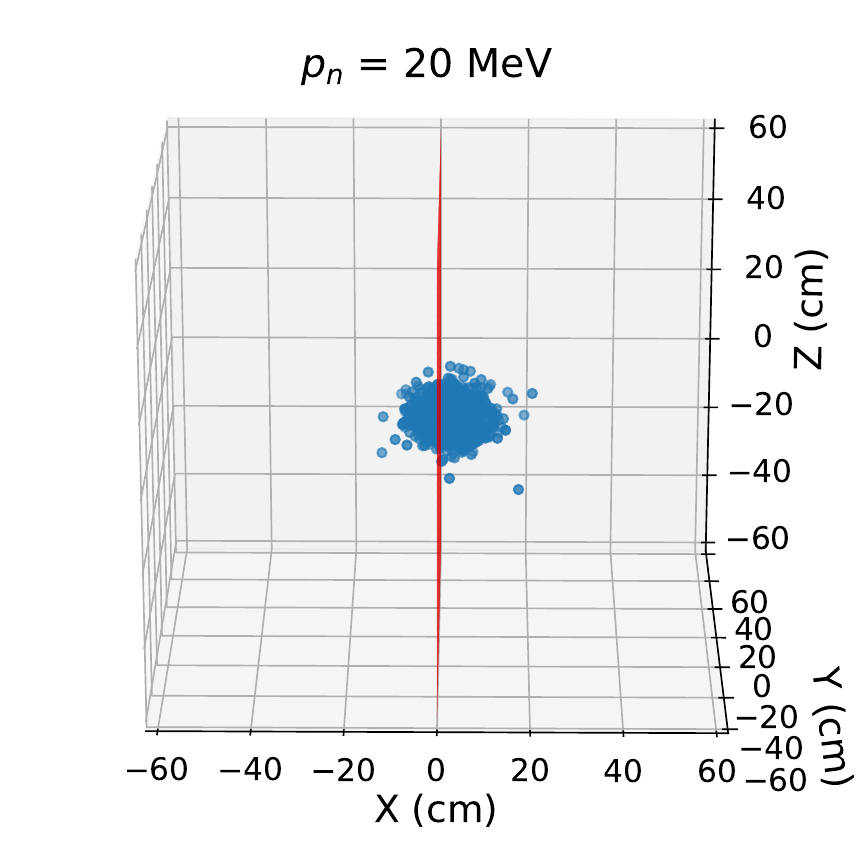}
    \label{fig:image1}
  \end{minipage}
  \hfill
  \begin{minipage}[b]{0.325\textwidth}
    \includegraphics[width=\textwidth]{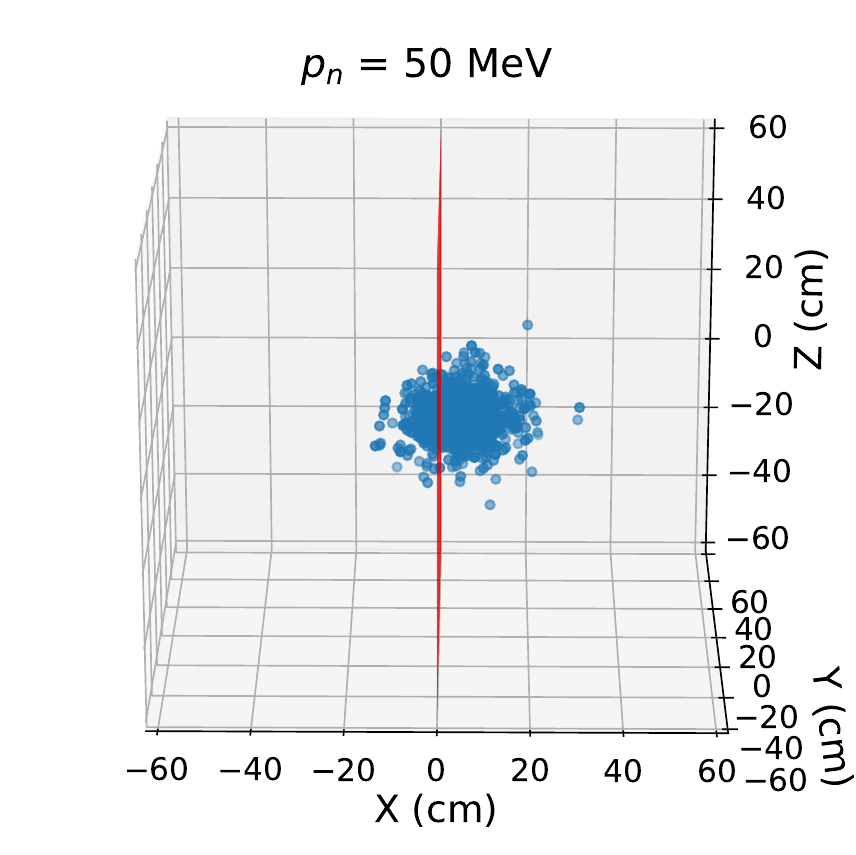}
    \label{fig:image2}
  \end{minipage}
  \hfill
  \begin{minipage}[b]{0.325\textwidth}
    \includegraphics[width=\textwidth]{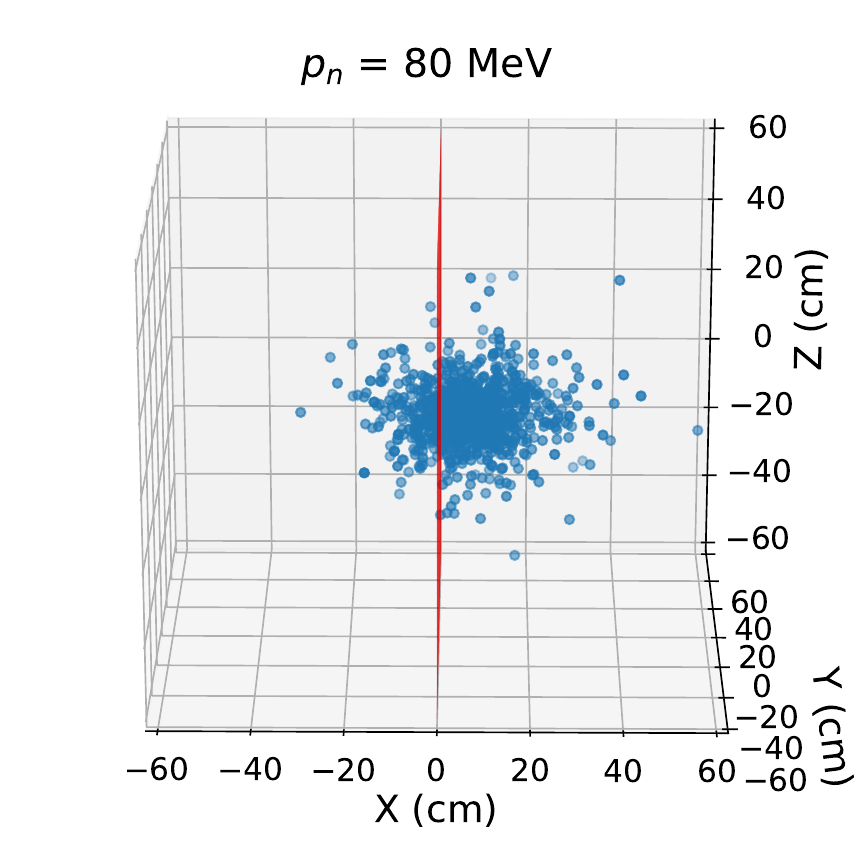}
    \label{fig:image3}
  \end{minipage}
  \caption{Neutron capture positions when 1000 neutrons with varying momenta of 20~MeV (left), 50~MeV (middle) and 80~MeV (right) are injected along the x-axis in Gd-loaded water.}
  \label{fig:three_images}
\end{figure*}

\subsection{Neutron Propagation and Capture Simulation}
We simulate the diffusion of neutrons in both pure water and Gd-doped water by using Geant4-v11.1.1~\cite{Allison:2016lfl,GEANT4:2002zbu}.

Geant4 is a comprehensive software toolkit for simulating the passage of particles through matter. It employs Monte Carlo methods to model particle interactions with materials, tracking their trajectories and simulating the effects of both electromagnetic and nuclear processes~\cite{Allison:2016lfl,GEANT4:2002zbu}. The Physics list in Geant4 is a collection of modules that can be combined to simulate various physical processes.
In this project, it is essential to include two specific modules: \texttt{G4HadronElasticPhysicsHP}, which handles hadron nuclear elastic processes, and \texttt{G4HadronPhysicsFTFP BERT HP}, which incorporates high precision neutron models for neutron capture. The Geant4 simulation overestimated the thermal motion of hydrogen atoms, causing an underestimation of hydrogen capture compared to Gd~\cite{Hino:2024mbb}. However, at a high Gd concentration (0.1\% Gd), this discrepancy is not significant.

In the simulation configuration, neutrons are injected at the origin in the positive $x$ direction. We consider two containers: one is pure water and one is 0.1\% concentration Gd-loaded water. 
To gather event information during the simulation, the stepping function is used. This function allows for the collection of energy deposition and distance traveled step by step. Moreover, it provides access to the process name and  the identities of the target particles involved in each step. 

Figure~\ref{fig:three_images} shows the neutron capture positions in Gd-loaded water for 20 MeV, 50 MeV, and 80 MeV neutron momenta, respectively. 
We find that more neutrons are captured in the positive $x$ direction compared to the negative direction in Gd-loaded water, and the asymmetry is larger for larger neutron momenta.  
Thus, depending on the neutron momentum, some directionality is indeed retained, which makes it theoretically possible to use the neutron capture information to do event reconstruction.

\begin{figure}[htbp]
    \centering
    \includegraphics[width=3.50in,height=3in]{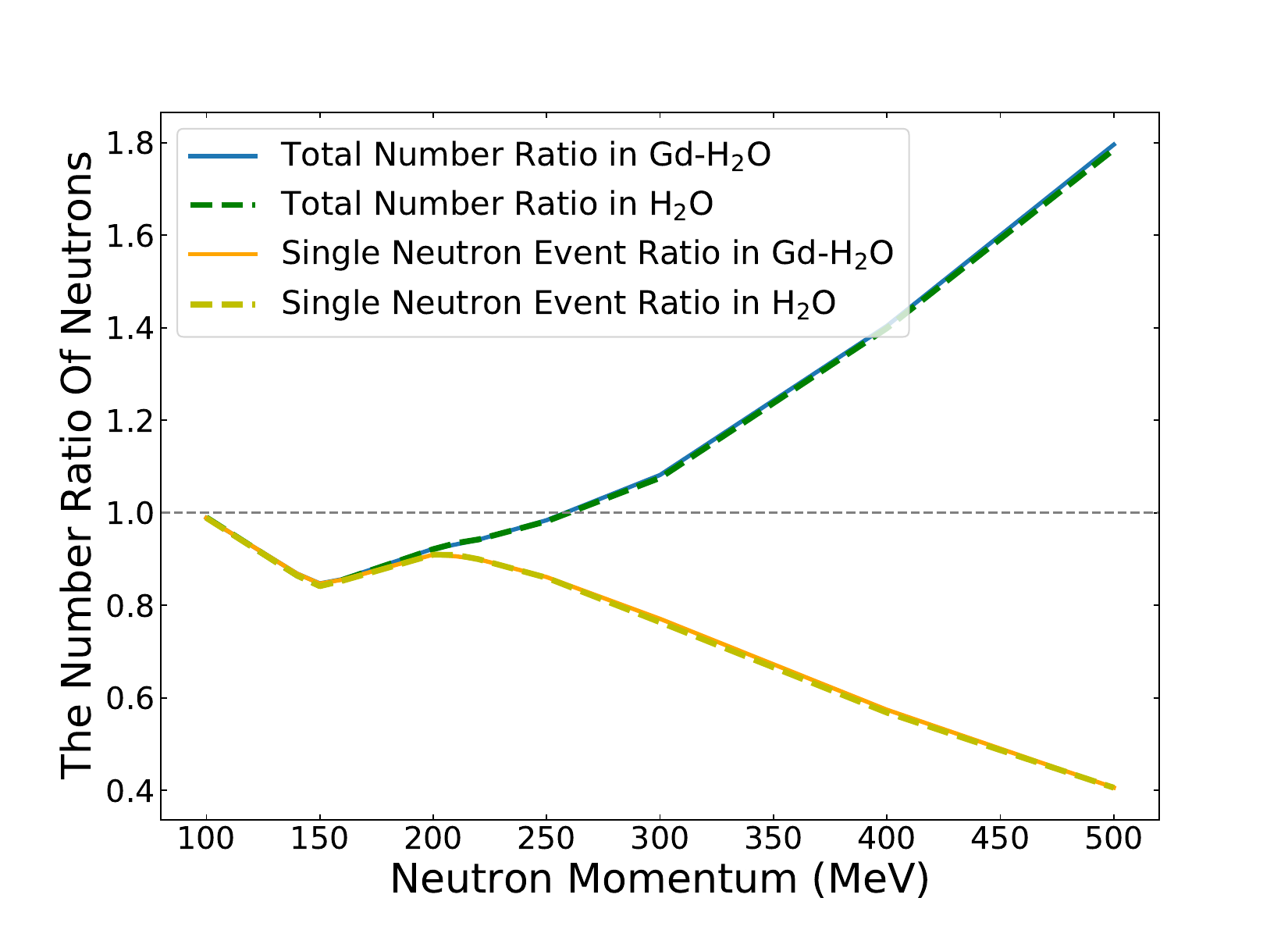}
    \caption{The total number ratio of neutron capture events with the emitted neutron is represented by the blue solid line~(Gd-loaded water) and the green dashed line~(pure water). The ratio of single neutron events is depicted by the orange solid line~(Gd-loaded water) and the yellow dashed line~(pure water).}
    \label{fig:rsele}
\end{figure}

Figure~\ref{fig:rsele} shows the number of neutron capture events per neutron injection. 
Below 100 MeV neutron momentum, this ratio is 1.  However, we find that the ratio decreases above 100~MeV, and then increases above 150~MeV.  This is because above 100~MeV, neutrons undergo inelastic scattering with $ ^{16}\text{O} $. This process can result in the transmutation of $ ^{16}\text{O} $ into other isotopes such as $ ^{15}\text{O} $ or $ ^{14}\text{O} $, or even lead to the formation of different elements such as $ ^{16}\text{N} $ and $ ^{13}\text{C} $. These interactions can either eliminate existing neutrons or produce new ones.  At higher energies, inelastic interactions become more and more important, and produce more neutrons, resulting in the increasing ratio.

In this work, we focus on events with single neutron capture, which we believe would be easy to identify and analyze experimentally; we defer the consideration of multiple neutron events to future work.  Imposing a cut to only single neutron events, we observe a dip at 150 MeV by the orange line and yellow dashed line in figure~\ref{fig:rsele}, due to the reasons discussed above.  
We find that at 500\,MeV, about 40\% of the events are single neutron events, which roughly sets the high-energy limit of consideration in this study.

\subsection{Neutron Capture Information}
Neutron capture events can provide several key pieces of information: capture time, capture distance, and capture angle.  While the capture time and distance are detailed in Appendix~\ref{appendix:ncapture},  we focus on the capture angle in this section.

\begin{figure}[htbp]
    \centering
    \includegraphics[width=3.5in,height=2.9in]{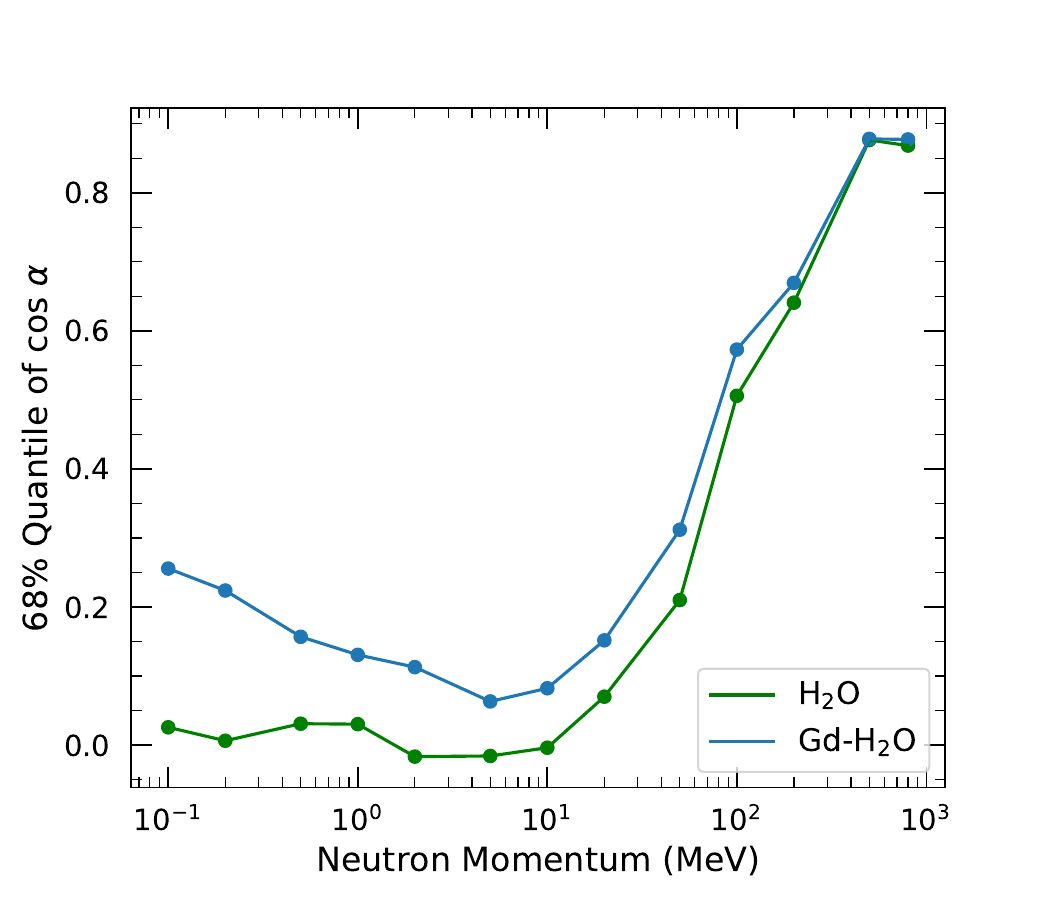}
    \caption{The neutron angular resolution as  a  function  of  neutron momentum, with the angular resolution defined as the 68\% quantile from $\cos\alpha = 1$. 
    }
    \label{fig:ar}
\end{figure}

Figure~\ref{fig:ar} shows the neutron ``angular resolution" as a function of the neutron momentum in both water and Gd-loaded water.  Here, we define the ``angular resolution" as the 68\% quantile from $\cos\alpha = 1$, where $\cos\alpha$ is the dot product between the final neutron capture position unit vector and the initial neutron momentum unit vector. 

As shown, neutrons statistically retain directional information even after diffusion in water, with a slight improvement in resolution observed in Gd-loaded water. Notably, the results indicate that even at low momentum, there is still a discernible directionality. Above 10\,MeV, the resolution improves dramatically, meaning that there is a better correlation between the final neutron capture position and the initial neutron momentum.  In this work, we focus on using the final neutron position for the event reconstruction. 

\section{Reconstruction of Neutrino Direction in IBD}
In this section, we explore the use of neutron capture information to reconstruct neutrino direction, specifically, with the neutron capture angle.  We first consider the ideal case of accounting only for neutron diffusion. Then we further investigate the impact of neutron capture and IBD interaction vertex resolutions, positron angular resolution, and positron energy resolution. Finally, we incorporate all of these factors into our study.

As a proof of principle study, we do not consider any  angular or energy distributions from real neutrino sources.  Instead, we assume a point source with a monochromatic energy distribution, and investigate how much angular information could be inferred using IBD event reconstruction including the neutron capture information. Given that the neutron tagging efficiency in Gd-loaded water is significantly higher than in pure water, our analysis primarily focuses on Gd-loaded water, with a discussion on pure water at the end. 

\subsection{Reconstruction of Neutrino Direction Considering Only Diffusion}

We begin by fixing the neutrino energy \( E_\nu \), which also implicitly sets the neutrino direction, corresponding to a point source with a monochromatic energy distribution. For a given \( E_\nu \), we consider 20 positron energy values, linearly spaced between the kinematic maximum and minimum. 
The contribution of each positron energy bin is weighted by the differential cross section, which determines the probability distribution of positron energies, as detailed in Appendix~\ref{appendix:xsection}.  Given \( E_\nu \) and the measurable \( E_e \), all remaining kinematic quantities (e.g., neutron momentum $p_n$, scattering angles $\phi, \theta$) are uniquely determined via energy momentum conservation in Eq.~(\ref{eq:ibd}). 
We note that even though we have implicitly assumed a fixed direction for the neutrinos, different neutrino angle solutions, \( \phi \), are obtained. This is because the reference framework is defined based on the positron momentum, as shown in Fig.~\ref{fig:ibd}.

For each positron energy bin, we then match the corresponding neutron momentum to the precomputed diffusion simulations. Neutron diffusion simulations are performed for neutron momenta from 1 to 800 MeV at intervals of 1 MeV, with each momentum bin containing $10^4$ events, and the neutron capture positions of single neutron capture events are recorded. These precomputed datasets form the discrete simulation samples used for subsequent neutron momentum assignment.

These capture positions are used to calculate the neutron capture angle $\theta'$ relative to the positron direction. Due to diffusion, the original angle $\theta$ transforms into a new angle, $\theta'$, as shown in Fig.~\ref{fig:ibd}. For each measurable experimental pair $\theta'$ and the positron energy $E_e$, we can reconstruct the neutron momentum, the neutrino energy and the neutrino direction $\phi'$ based on Eq.~(\ref{eq:ibd}), with details and an example of averaged positron energy utilization given in Appendix~\ref{appendix:positron}. Finally, we  compare $\phi'$ to the original direction $\phi $ by considering $\cos \psi= \cos (\phi' - \phi)$, with $\cos \psi$ closer to 1 representing better accuracy.

The theoretical reconstruction uncertainties from the IBD cross section primarily arise from the form factor uncertainties discussed in Appendix~\ref{appendix:form}. These contributions are negligible compared to detector effects.

\begin{figure}[htbp]
    \centering
    \includegraphics[width=3.50in,height=3in]{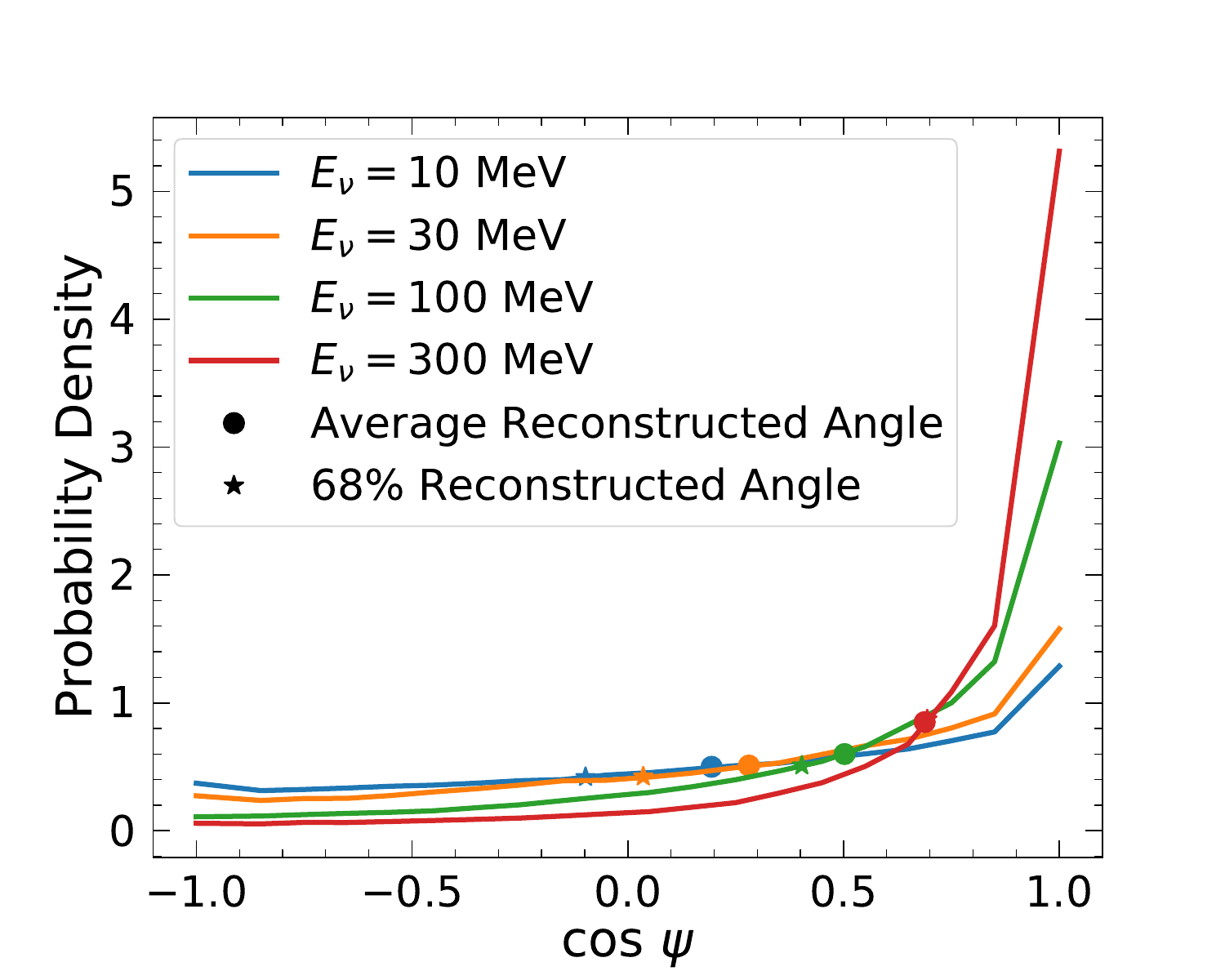}
    \caption{The reconstruction of neutrino direction resolution for neutrino energies of 10, 30, 100, and 300 MeV. The 68\% quantile from $\cos\psi = 1$ is indicated by stars, while the average reconstructed angle is represented by circles. The red star and red circle overlap each other.}
    \label{fig:30crs4}
\end{figure}

Figure~\ref{fig:30crs4} shows the probability distribution of $\cos\psi$ for several neutrino energies. Even with neutron diffusion, the inclusion of neutron capture information demonstrably enhances the reconstruction of neutrino directionality. For the case of 30 MeV neutrino energy, the average $\cos\psi$ is about 0.28, which provides significantly improved angular resolution compared to relying solely on positron information. As discussed in detail in the Appendix~\ref{appendix:xsection} and illustrated in Fig.~\ref{fig:30p}, the positron angular distribution is governed by the cross section, and for 30 MeV neutrinos this distribution is essentially flat.

\subsection{Detector Resolutions}
In practical applications, the consideration of detector vertex resolution is crucial. References~\cite{Super-Kamiokande:2008ecj,Super-Kamiokande:2010tar,Super-Kamiokande:2023jbt,Super-Kamiokande:2016yck} provide information on the vertex, angular, and energy resolutions for each phase of the SK experiment. The significant improvement in SK-III compared to SK-I is due to the implementation of an advanced vertex reconstruction program~\cite{Super-Kamiokande:2023jbt}. Further enhancements observed in SK-IV are attributed to improved timing resolution and a better match between the timing residuals of data and Monte Carlo simulated events~\cite{Super-Kamiokande:2023jbt}. A reconstruction algorithm designed to enhance vertex resolution for positrons and neutrons has been developed~\cite{Kneale:2022sht}. However, current vertex resolution remains worse than the neutron diffusion distance at lower neutron momentum.

In the SK-IV experiment, the vertex resolution for electrons in the energy range from a few to tens of MeV is approximately 50 cm, improving to 30 cm at higher energies~\cite{Super-Kamiokande:2023jbt}. The angular resolution for electrons with energies above 30 MeV can reach approximately 10$^\circ$, while the energy resolution achieves around 10\% for electrons in the tens of MeV range~\cite{Super-Kamiokande:2023jbt}. For neutron capture vertices in SK, the reconstruction resolution is approximately 55 cm for captures on Gd and 110 cm for captures on hydrogen~\cite{Abe:2024ydm}. In this study, we adopt 30 cm as the baseline vertex resolution, corresponding to the best-case value achieved by current water Cherenkov detectors at high energies, and additionally consider 10 cm as a target benchmark for future reconstruction algorithms or for small detectors which achieve better vertex resolutions~\cite{Kneale:2022sht}. Both current neutron capture vertex resolutions lie on the poorer side of this baseline.

The detector resolutions for positron angular and energy are detailed in Appendix~\ref{appendix:positronresolution}. Here we focus on the vertex resolution.

\begin{table*}[htbp]
\centering
\begin{tabular}{|c|c|c|c|c|}
\hline
\diagbox{$E_\nu$}{$\cos\psi_{68\%}$} & $\delta d$=0~cm & $\delta d$=10~cm & $\delta d$=30~cm \\
\hline
10 MeV & -0.097 & -0.28  & -0.34 \\
\hline
30 MeV& 0.036 & -0.16  & -0.27 \\
\hline
100 MeV& 0.40 & 0.26 &  0.044 \\
\hline
300 MeV& 0.69 & 0.62 &  0.49 \\
\hline
\end{tabular}
\caption{The 68\% quantile neutrino direction reconstruction resolutions for varying neutrino energies, considering vertex resolutions of 10 cm and 30 cm.}
    \label{tab:verd}
\end{table*}

To account for the vertex resolution effect, we use a normal distribution to simulate the detected neutron capture position. We assume that the detected capture position is a 3-dimensional random point $\mathbf{x} = (x, y, z)$ with a normal distribution centered at the original capture position $\boldsymbol{\mu} = (\mu_x, \mu_y, \mu_z)$.
For each coordinate $x_i$ (where $i \in \{x, y, z\}$), we model the distribution using a normal probability density function:
\begin{equation}
\label{eq:normal}
f(x_i) = \frac{1}{\sigma \sqrt{2\pi}} e^{-\frac{(x_i - \mu_i)^2}{2\sigma^2}},
\end{equation}
where $\sigma$ is the standard deviation of the distribution, which is calculated as:
\begin{equation}
\sigma = \frac{\delta d}{\chi^2_{0.68, 3}},
\end{equation}
where $\delta d$ is the vertex resolution, and $\chi^2_{0.68, 3}$ is the chi-squared value for a 68\% probability with 3 degrees of freedom. This scaling ensures that 68\% of the simulated points lie within the desired distance from the mean. In this study, the same vertex resolution $\delta d$ is applied to both the IBD interaction vertex and the neutron capture vertex.
The chi-squared value can be calculated using the `scipy.stats.chi' module in Python,
chi.ppf(0.68, 3).

\begin{figure*}[htbp]
  \centering
  \begin{minipage}[b]{0.325\textwidth}
    \includegraphics[width=\textwidth]{figures/50cmn0cm0.pdf}
    \label{fig:image1r}
  \end{minipage}
  \hfill
  \begin{minipage}[b]{0.325\textwidth}
    \includegraphics[width=\textwidth]{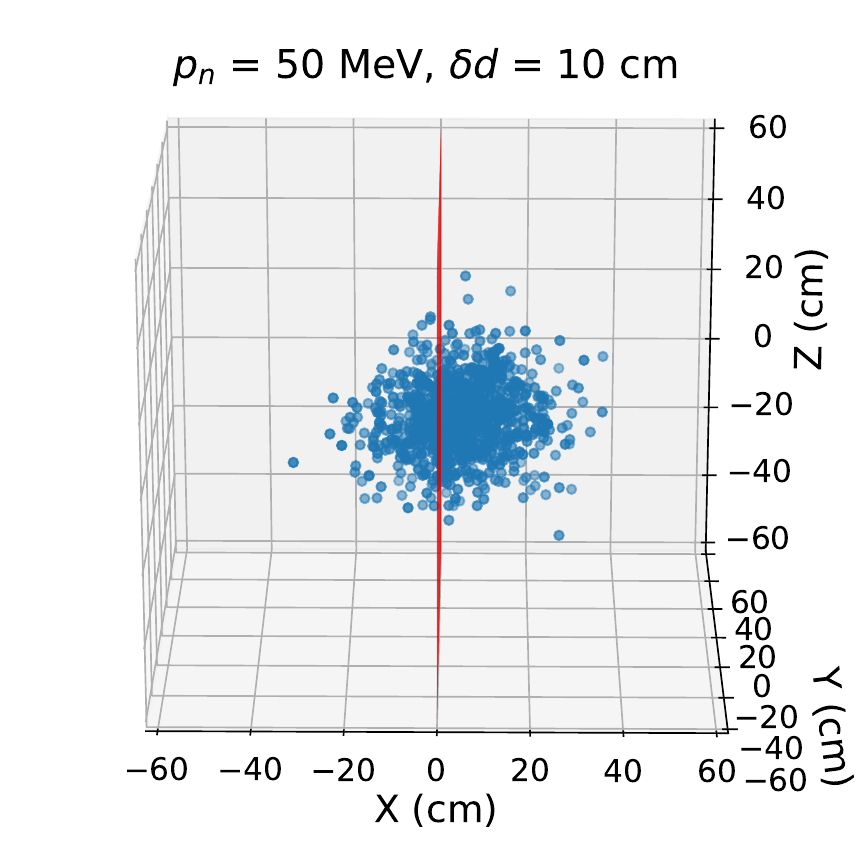}
    \label{fig:image2r}
  \end{minipage}
  \hfill
  \begin{minipage}[b]{0.325\textwidth}
    \includegraphics[width=\textwidth]{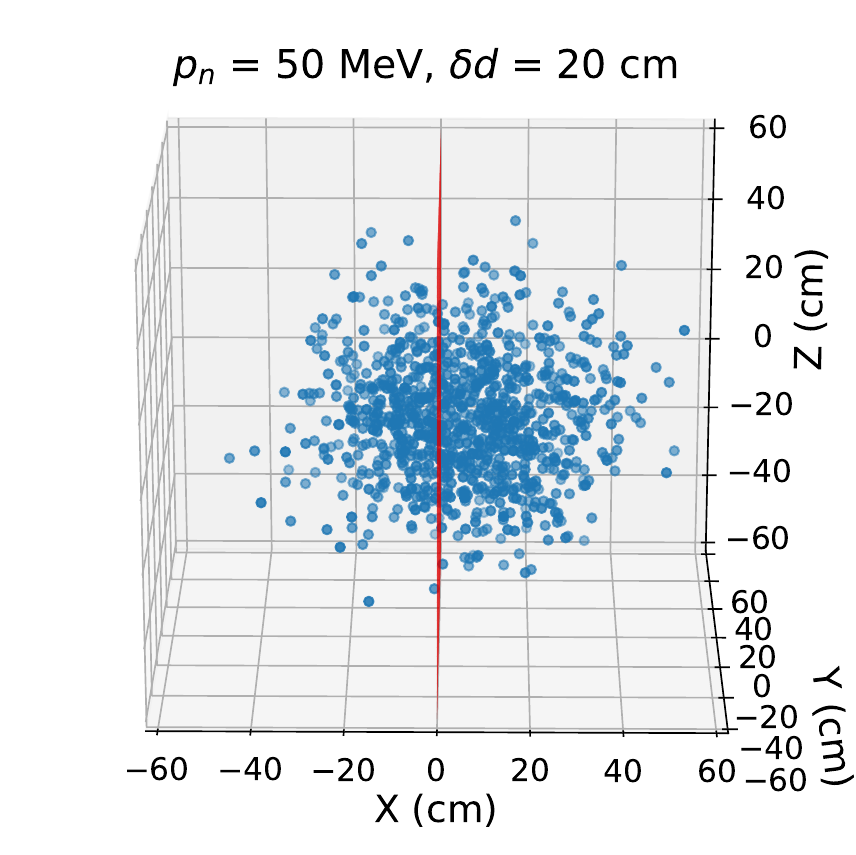}
    \label{fig:image3r}
  \end{minipage}
  
  \caption{50 MeV momentum neutron capture positions for varying vertex resolution of ideal case (left), 10~cm (middle) and 20~cm (right).The ideal case corresponds to the middle plot shown in Fig.~\ref{fig:three_images} above.}
  \label{fig:three_images_resolution}
\end{figure*}

Fig.~\ref{fig:three_images_resolution} shows the neutron capture positions with different vertex resolutions. We simulate 1000 neutron captures with 50 MeV momentum, applying Eq.~\ref{eq:normal} to incorporate vertex resolution effects for each capture point.
The resulting spatial distribution of neutron capture positions exhibits an asymmetry, with the number of capture positions at $x > 0$ consistently exceeding those at $x < 0$ across the vertex resolutions considered. This indicates that the neutron capture position could retain some of the original neutron directional information, even with some smearing effect caused by the vertex resolutions.

To study the effect of the vertex resolution on the reconstruction of the neutron direction, we perform the following simulation. For each neutron capture position ($10^4$ realizations per positron energy, as discussed above), using the numpy.random.normal function, we generate a new point with a Gaussian distribution centered at $(0,0,0)$ and a standard deviation $\delta d$, representing the neutron emission position adjusted for vertex resolution. Additionally, we generate another point with a Gaussian distribution centered at the simulated neutron capture position, also with a standard deviation of $\delta d$, to model the measured neutron capture position. The neutron capture angle $\theta'$ is then adjusted to $\theta''$. We use the new angle $\theta''$ along with the positron energy $E_e$ to perform the reconstruction.

Table~\ref{tab:verd} presents the 68\% quantile neutrino direction reconstruction resolutions $\cos\psi$ with vertex resolutions of 0 cm, 10 cm and 30 cm at various neutrino energies. Vertex resolutions have a substantial impact on the accuracy of neutrino direction reconstruction, highlighting the need for enhancements in vertex reconstruction for improved results. With $10^4$ neutron events per momentum bin in the precomputed diffusion samples, the statistical uncertainty of the reconstructed $\cos\psi_{68\%}$ is below 0.02, about an order of magnitude smaller than the resolution effects studied here, and is therefore neglected.

\subsection{Reconstruction with All Detector Effects}

\begin{figure*}[htbp]
  \centering
  
    \includegraphics[width=\textwidth]{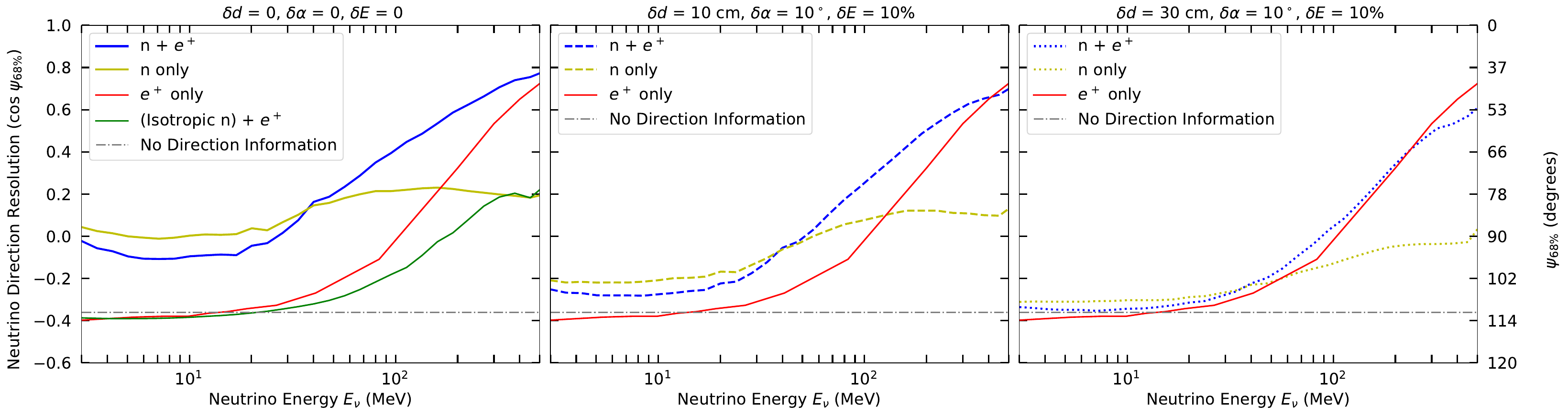}  
  \caption{The 68\% quantile of the neutrino direction reconstruction resolution is shown as a function of neutrino energy from 3 MeV to 500 MeV, for different resolution scenarios. In the left plot, the blue solid curve shows the reconstruction using positron and neutron diffusion information with perfect detector resolution. The yellow curve represents experiments considering only the neutron capture position, excluding positron data. The red solid curve depicts the performance using only the positron direction, without accounting for any detector resolutions. The green curve follows the same method as the blue curve but assumes neutrons are emitted isotropically. The gray dot-dashed line indicates the 68\% quantile for the scenario without any directional information. The middle and right plots incorporate additional detector resolutions: a neutron capture and IBD vertex resolution of 10 cm and 30 cm, respectively, a positron angular resolution of 10$^\circ$, and a positron energy resolution of 10\%.}
  \label{fig:300recq}
\end{figure*}

In this section, we account for the vertex resolution, positron angular resolution, and positron energy resolution discussed in Appendix~\ref{appendix:positronresolution}. For each neutron capture event, utilizing the updated positron energy $E_e'$ and the recalculated neutron angle $\theta''$~(considering both vertex and positron angular resolutions), we apply the complete neutrino direction reconstruction algorithm as described in the previous section.

Figure~\ref{fig:300recq} shows the 68\% quantile of neutrino direction resolution for neutrino energies from 3 MeV to 500 MeV. The left panel shows the reconstruction without detector resolution effects, while the middle and right panels include additional detector resolutions, assuming a positron angular resolution of 10$^\circ$ and a positron energy resolution of 10\%. The dashed and dotted curves correspond to neutron capture vertex and IBD vertex resolutions of 10 cm (middle panel) and 30 cm (right panel), respectively. The blue solid curve represents the neutrino direction resolution achieved by incorporating both positron direction and neutron diffusion information, assuming ideal detector performance. In contrast, the red solid curve depicts the resolution when relying solely on positron direction (neutrino direction = positron direction), also assuming ideal detector performance.
The dot-dashed grey line corresponds to the case of no directional information. The red curve falls below this line at low energies, indicating a slight backward bias in positron emission direction relative to the neutrino direction~\cite{Vogel:1999zy}.

Across all energies, the blue solid curve consistently outperforms the red curve, demonstrating the significant improvement in IBD angular resolution gained by incorporating neutron capture information. In comparing the three panels, the reconstruction resolution exhibits negligible improvement with a 30 cm vertex resolution. A vertex resolution of 30 cm already provides little improvement over the positron-only reconstruction, and at neutrino energies below 10 MeV essentially no directional information is retained. Any vertex resolution worse than 30 cm is therefore insufficient for the method to be effective. However, if one can improve the vertex resolution to 10 cm, it could substantially enhance neutrino direction reconstruction.

The average neutrino direction resolution is provided in Fig.~\ref{fig:300reca} in Appendix~\ref{appendix:resolution}.
\subsection{Reconstruction with Isotropic Neutron and Reconstruction with Neutron Direction Only}

\begin{figure}[htbp]
    \centering
    \includegraphics[width=3.60in,height=2.7in]{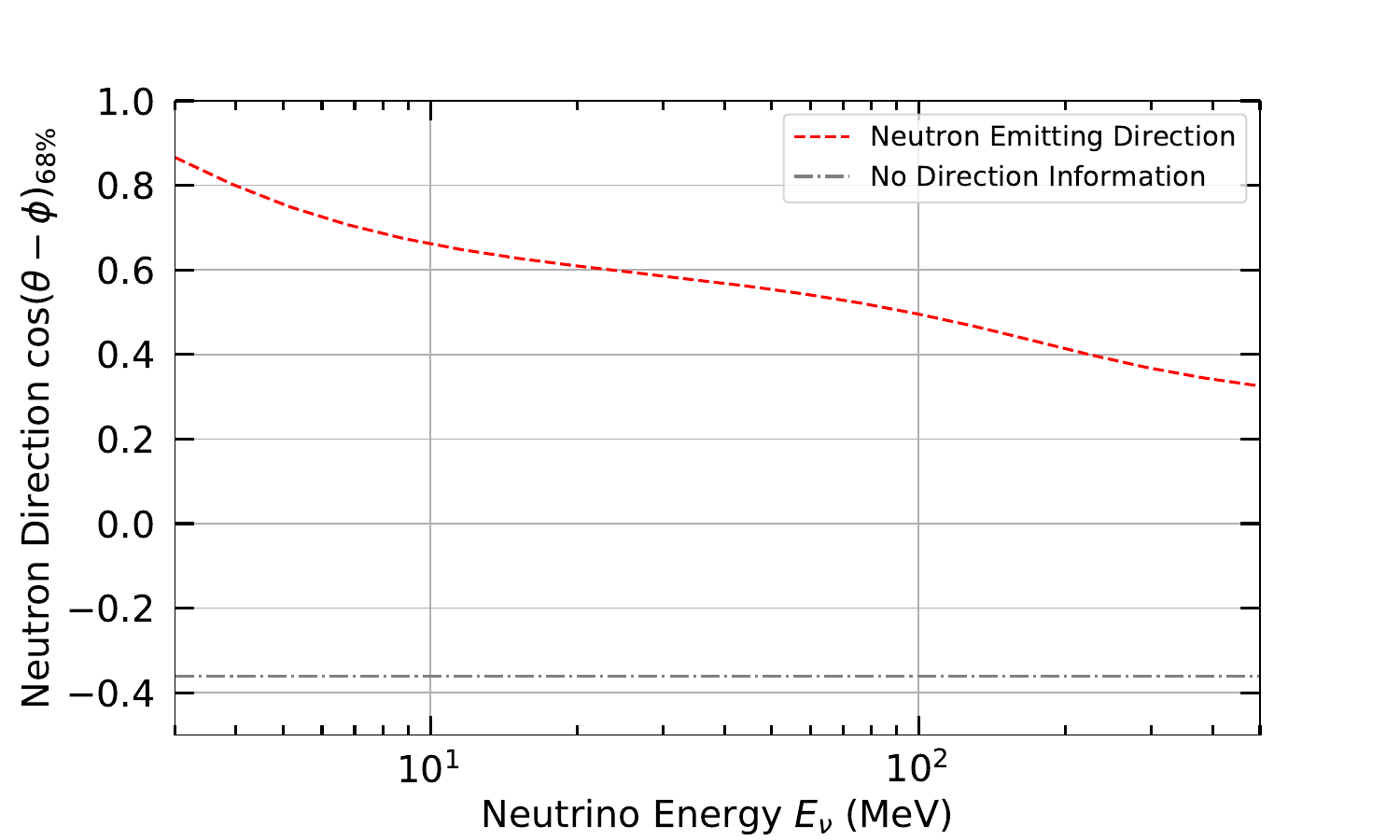}
    \caption{ The 68\% quantile of neutron emitting direction with respect to neutrino direction for neutrino energies from 3 MeV to 500 MeV is shown. }
    \label{fig:500n}
\end{figure}

We also examine the scenario where the outgoing neutron direction is entirely isotropic, which corresponds to situations with very poor detector resolution. To simulate this, we create a set of isotropic neutron capture positions and perform the reconstruction. The resulting 68\% quantile of the neutrino direction resolution is depicted by the green line in Fig.~\ref{fig:300recq} (left).
For low-energy neutrinos, both the positron and neutron are isotropic, resulting in an almost isotropic reconstruction of the neutrino direction. For high-energy neutrinos, while positrons carry some directional information, the isotropic neutrons reduce this directional information, thus the directionality is worse than the red solid line.

We now consider a scenario where only neutron capture position information is utilized (neutrino direction = neutron direction), incorporating neutron diffusion effects while excluding positron direction data. The results, shown as yellow lines in Fig.~\ref{fig:300recq}, demonstrate that for low-energy neutrinos (below 40 MeV), this neutron-only approach surpasses the performance of the combined method~($n+e^+$). This advantage arises because the isotropic emission of positrons at these energies diminishes the effectiveness of directional information in the combined method. Consequently, neutron-only data can be effectively used to reconstruct the neutrino direction for energies below 40 MeV.
This can be further illustrated in Fig.~\ref{fig:500n}, where we show the 68\% quantile of the neutron emission direction relative to the neutrino direction, according to the differential cross section. We can see that neutrons are preferentially emitted along the neutrino direction, especially at lower energies.

At higher neutrino energies, the directional information conveyed by the neutrons is limited by their increasing emission angle deviation from the neutrino direction. As the neutrino energy rises, the alignment between neutron emission and neutrino directions weakens, reducing the overall directional resolution achievable with neutron data alone. Consequently, at higher energies, the positron direction becomes more critical for achieving accurate neutrino direction reconstruction.

The neutron-only approach is particularly interesting to liquid scintillator detectors such as JUNO~\cite{JUNO:2015zny}, where distinguishing between Cherenkov and scintillation light remains a significant challenge, complicating the determination of the positron direction. However, JUNO’s superior vertex resolution enables precise reconstruction of neutron capture positions. The neutron-only approach 
in Gd-loaded water Cherenkov detectors is analogous to those proposed in liquid scintillator detectors~\cite{Vogel:1999zy,CHOOZ:1999hgz,Fischer:2015oma,Mukhopadhyay:2020ubs,Li:2020gaz}.

\subsection{Comparison of Reconstruction in Gd-loaded Water and Pure Water}
\begin{figure*}[htbp]
  \centering
  
    \includegraphics[width=\textwidth]{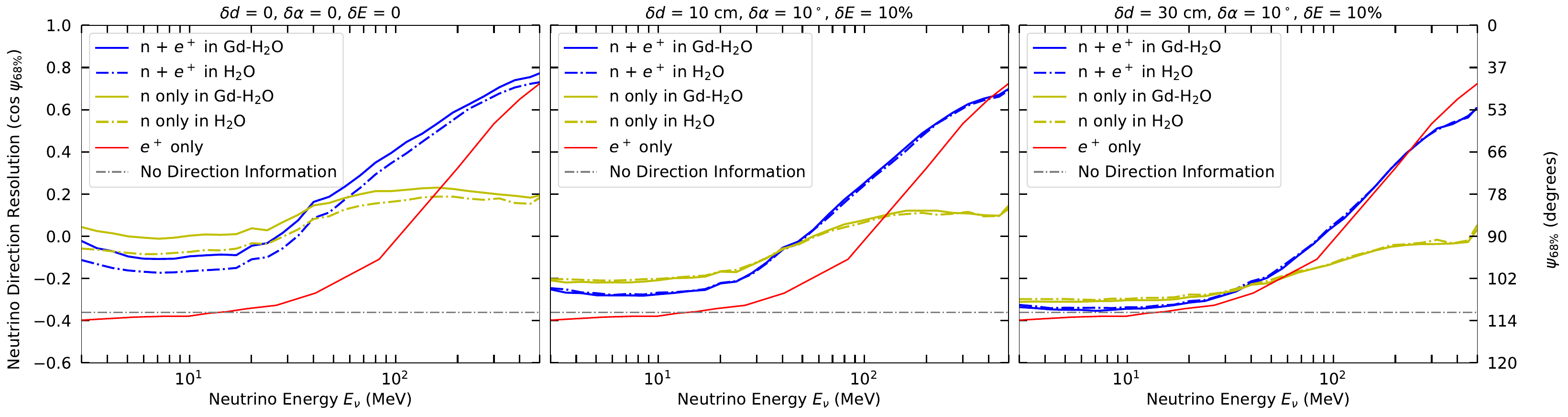}

\caption{The 68\% quantile of the neutrino direction reconstruction resolutions in both water and Gd-loaded water are shown as a function of neutrino energy from 3 MeV to 500 MeV, for different resolution scenarios. In the left plot, the blue curves show the reconstructions using positron and neutron diffusion information with perfect detector resolution in both water(dash-dot) and Gd-loaded water(solid). The yellow curves represent experiments considering only the neutron capture direction, excluding positron data. The red solid curve depicts the performance using only the positron direction. The middle and right plots incorporate additional experimental resolutions: a neutron capture and IBD vertex resolution of 10 cm and 30 cm, respectively, a positron angular resolution of 10$^\circ$, and a positron energy resolution of 10\%. The red solid curves in these plots are based solely on the positron direction, without accounting for any experimental resolutions. }
  \label{fig:300recqw}
\end{figure*}

Fig.~\ref{fig:300recqw} shows a comparison of the neutrino direction reconstruction between Gd-loaded water and pure water. The left plot shows that the reconstruction resolution is better in Gd-loaded water than in pure water. This is due to the fact that neutrons retain more directional information in Gd-loaded water than in pure water, as illustrated in Fig.~\ref{fig:ar}.
However, when considering the experimental resolutions, the reconstructions become comparable between Gd-loaded water and pure water. This is because the neutron capture distance in pure water is longer than in Gd-loaded water, as shown in Fig.~\ref{fig:acd}. The longer capture distance in pure water helps to compensate for the loss of directional information, leading to similar reconstruction performance. 
Notably, neutron tagging efficiency in pure water is lower than in Gd-loaded water~\cite{Beacom:2003nk}, and less light is produced after neutron capture, which may limit the practical effectiveness of this approach.

\section{Conclusions and Discussions}
In this study, we demonstrate a novel approach that could enhance the neutrino directionality of IBD events in water Cherenkov detectors, using the neutron capture information. Due to neutron diffusion in water, event-by-event reconstruction is not feasible. However, even with diffusion, we find that neutron capture information could significantly enhance the neutrino directionality statistically. 

For neutrinos with energies below 40 MeV, we find that the use of neutron capture position alone yields a superior angular reconstruction performance compared to the combined approach incorporating both neutron and positron information. This is because the isotropic nature of positrons tends to diminish the effectiveness of the directional information from the neutron at these lower energies.

However, the ability to recover neutrino direction depends crucially on the ability to determine the interaction vertex and the neutron capture position, which is used to find the neutron capture position vector. For a vertex resolution of 30 cm on both vertices, we find that the directionality is lost, and this approach offers no improvement over positron-only results. Current SK-Gd neutron capture vertex resolutions are larger than this baseline, confirming that the method is not yet effective with present detector performance. 
Our work then motivates consideration of methods to obtain better vertex information (e.g., Ref.~\cite{Kneale:2022sht}) either through better algorithms or detector modifications. Also, we note that it is not optimal to reconstruct two vertices and then find the vector between them. In practice, one can imagine it is better to have a single algorithm to directly reconstruct the neutron capture vector using the photon distributions. This may reduce the detector requirements needed for enhancing the neutrino directionality. 

Building on these considerations, our findings have important implications for next-generation neutrino detectors, particularly Hyper-Kamiokande (HK)~\cite{Hyper-Kamiokande:2018ofw}, which is expected to begin operation around 2028. The methodology developed here can be extended to HK, offering valuable guidance for optimizing its physics potential.

This improved directional information could be important for supernova pointing applications, improving early-warning capabilities~\cite{CHOOZ:1999hgz,Fischer:2015oma,Mukhopadhyay:2020ubs,Li:2020gaz}. Neutrinos are emitted during a supernova event before the optical light, allowing for a better early warning system. Detecting the direction of these neutrinos can quickly narrow down the search area in the sky, aiding in the identification of progenitor stars, even in cases of visually dim supernovae. 

Directional neutrino detection is also invaluable in the search for DM. Neutrinos could be produced through DM annihilation in celestial bodies such as the Sun or the Galactic Center.
Using directional information enhances the sensitivity of DM searches by narrowing down the search area in the sky and distinguishing potential DM signals from other neutrino sources and background noise, thereby giving a better constraint on the properties of DM. For a comprehensive review of DM annihilation into neutrinos, see~\cite{Arguelles:2019ouk}.

Furthermore, the application of directionality extends to non-proliferation and reactor monitoring~\cite{WATCHMAN:2015lcq}. Directional detection of neutrinos from reactors can help verify compliance with nuclear treaties by remotely monitoring reactor activity.

Our work serves as a first step toward exploiting neutron capture information for neutrino reconstruction in Gd-loaded water Cherenkov detectors, with SK-Gd taken as a benchmark configuration. We focus on neutrino directionality of IBD reactions, considering only the neutron capture angle relative to the positron vector. Other observables (e.g., neutron capture distance) and reactions beyond IBD may also be useful, but we leave these for future work.


\appendix
\section{Neutron Capture}
\label{appendix:ncapture}

\begin{figure*}[htbp]
  \centering
  \begin{minipage}[b]{0.45\textwidth}
    \includegraphics[width=\textwidth]{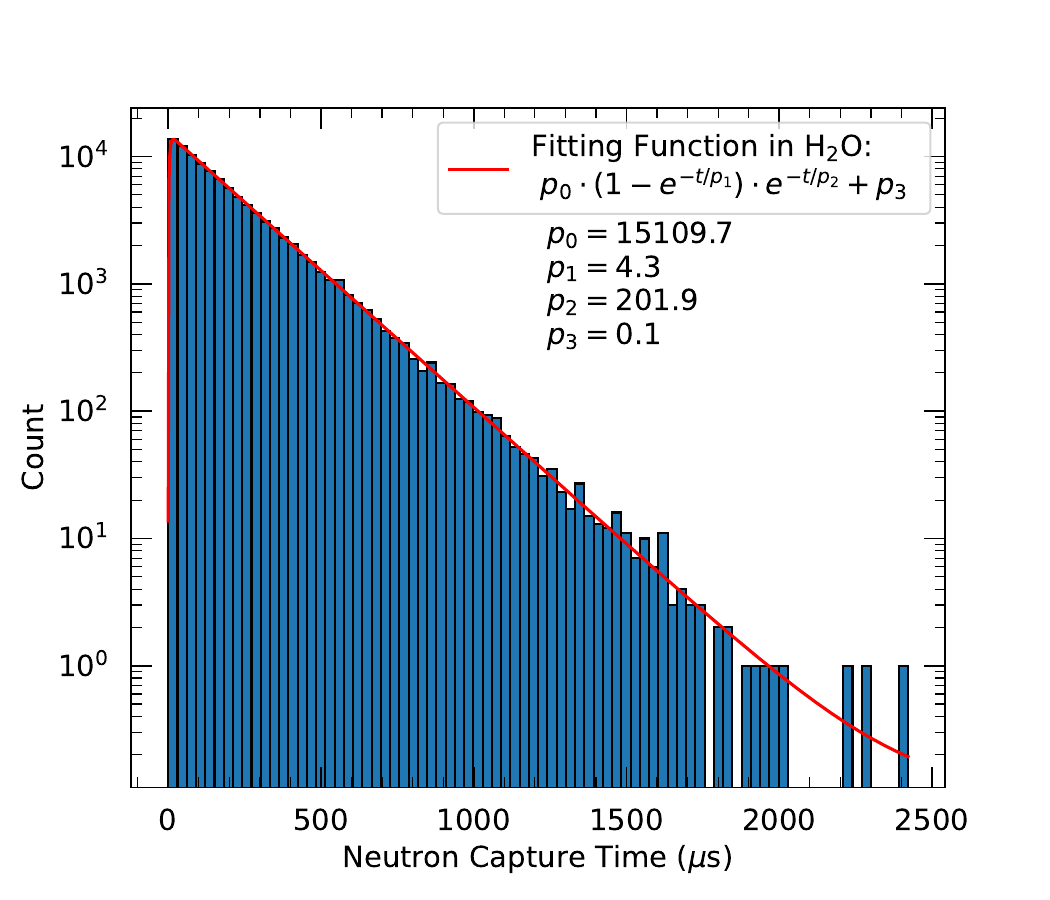}
    
    \label{fig:wt}
  \end{minipage}
  \hfill
  \begin{minipage}[b]{0.45\textwidth}
    \includegraphics[width=\textwidth]{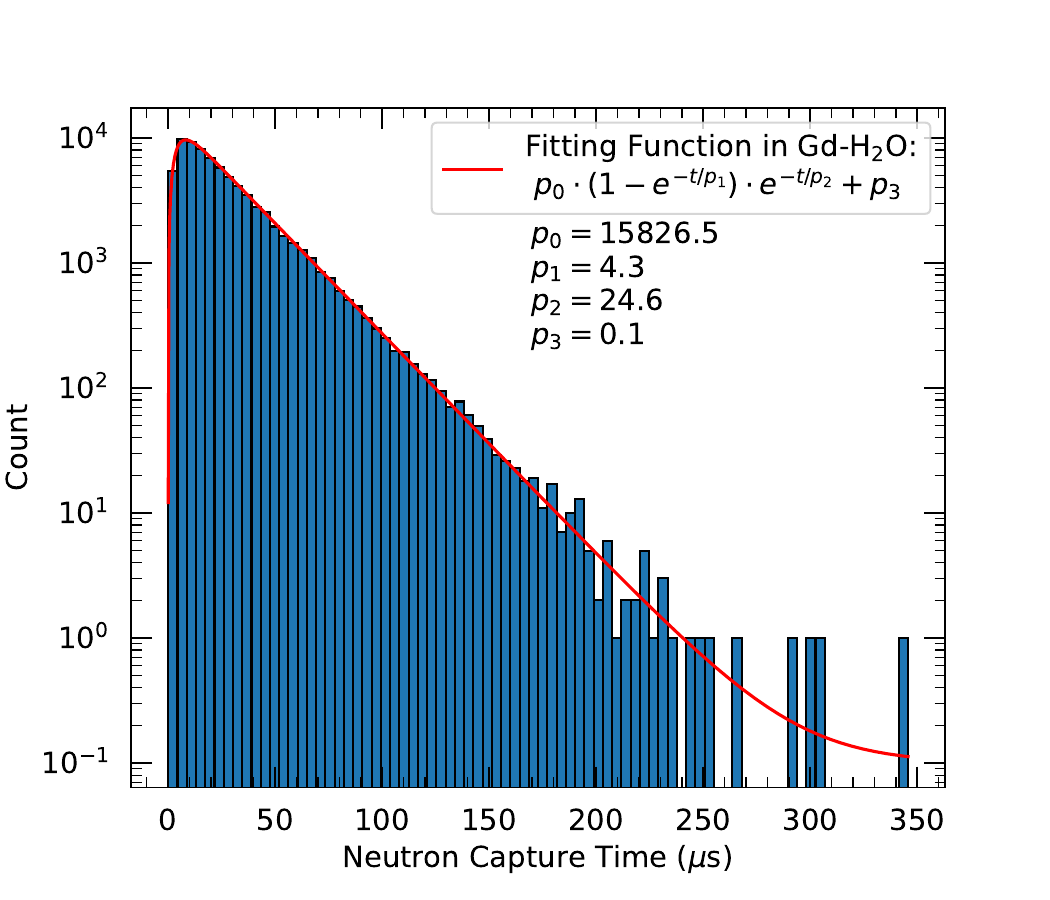}
    
    \label{fig:gdt}
  \end{minipage}
  \caption{In water (left) and Gd-loaded water (right), the capture times of $10^5$ neutrons with a momentum of 30 MeV are shown.}
  \label{fig:gdwt}
 
\end{figure*}

\begin{figure*}[htbp]
  \centering
  \begin{minipage}[b]{0.45\textwidth}
    \includegraphics[width=\textwidth]{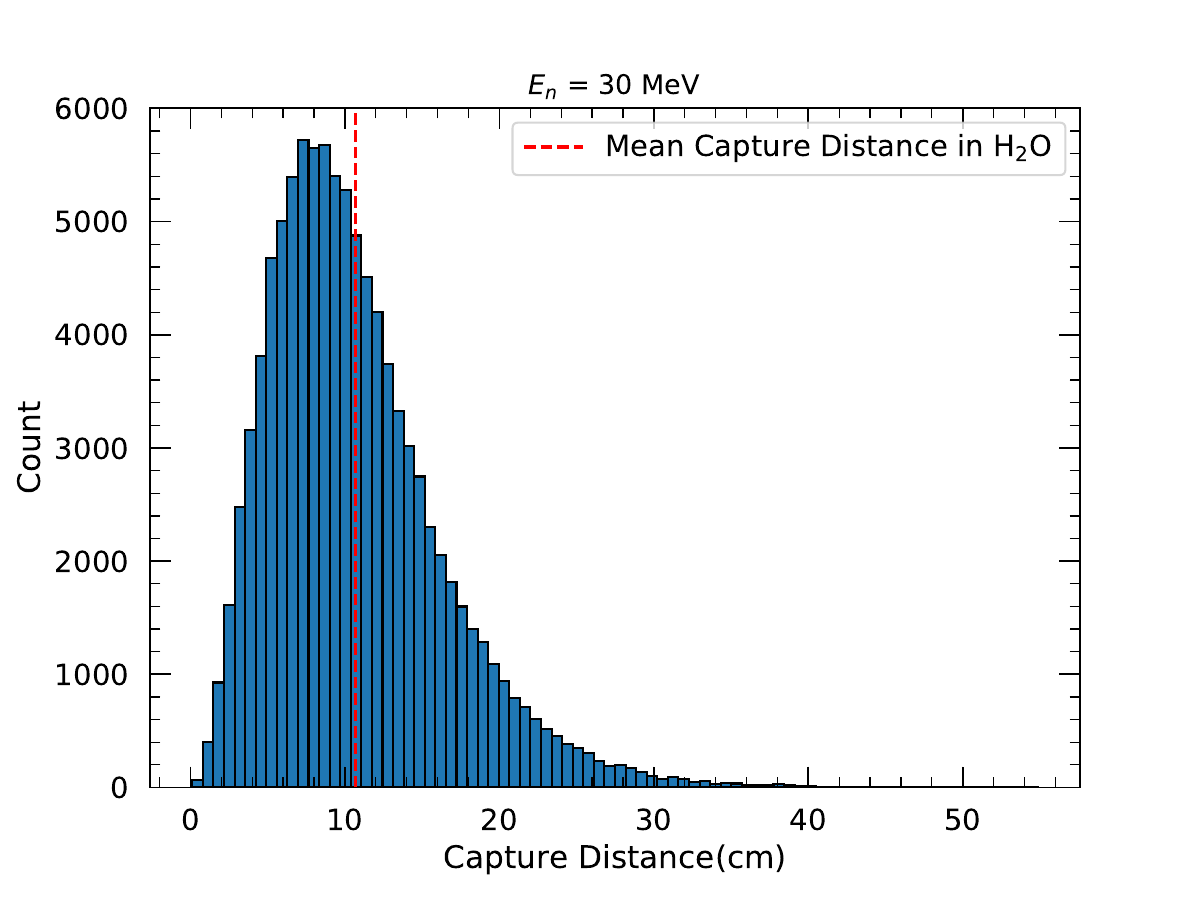}
    \label{fig:wd}
  \end{minipage}
  \hfill
  \begin{minipage}[b]{0.45\textwidth}
    \includegraphics[width=\textwidth]{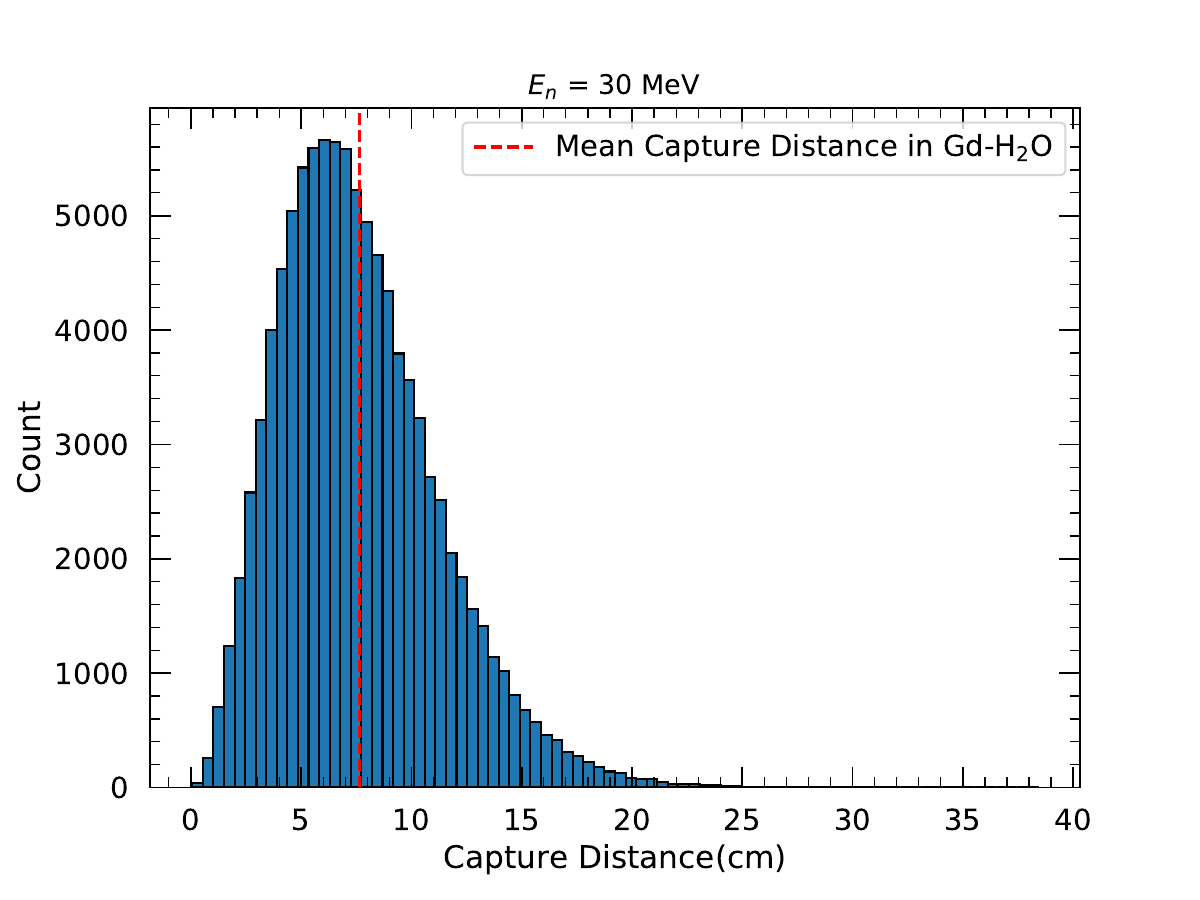}
    \label{fig:gdd}
  \end{minipage}
  \caption{ In water~(left) and Gd-loaded water~(right), the capture distances of $10^5$ neutrons with a momentum of 30 MeV are shown.}
  \label{fig:gdwd}
\end{figure*}

Neutron capture time and position can be obtained from the Geant4 simulation. Fig.~\ref{fig:gdwt} and Fig.~\ref{fig:gdwd} show the capture time and the capture distance of $10^5$ neutrons with a momentum of 30 MeV in both water and Gd-loaded water.
\subsection{Neutron Capture Time}
The fitting function for the neutron capture time is given by~\cite{Super-Kamiokande:2021the},
\begin{equation}
 f(t) = p_{0} \cdot (1 - e^{-\frac{t}{p_{1}}}) \cdot e^{-\frac{t}{p_{2}}} + p_{3},
\end{equation}
where the thermalization time constant \( p_{1} \), fixed to be 4.3 $\mu$s, is obtained from a combined analysis of the measurements~\cite{Super-Kamiokande:2021the}, and \( p_{0} \), \( p_{2} \), and \( p_{3} \) are fitting parameters. The parameter \( p_{2} \) represents the neutron capture time constant.

\begin{figure}[t]
    \centering
    \includegraphics[width=3.5in,height=3.5in]{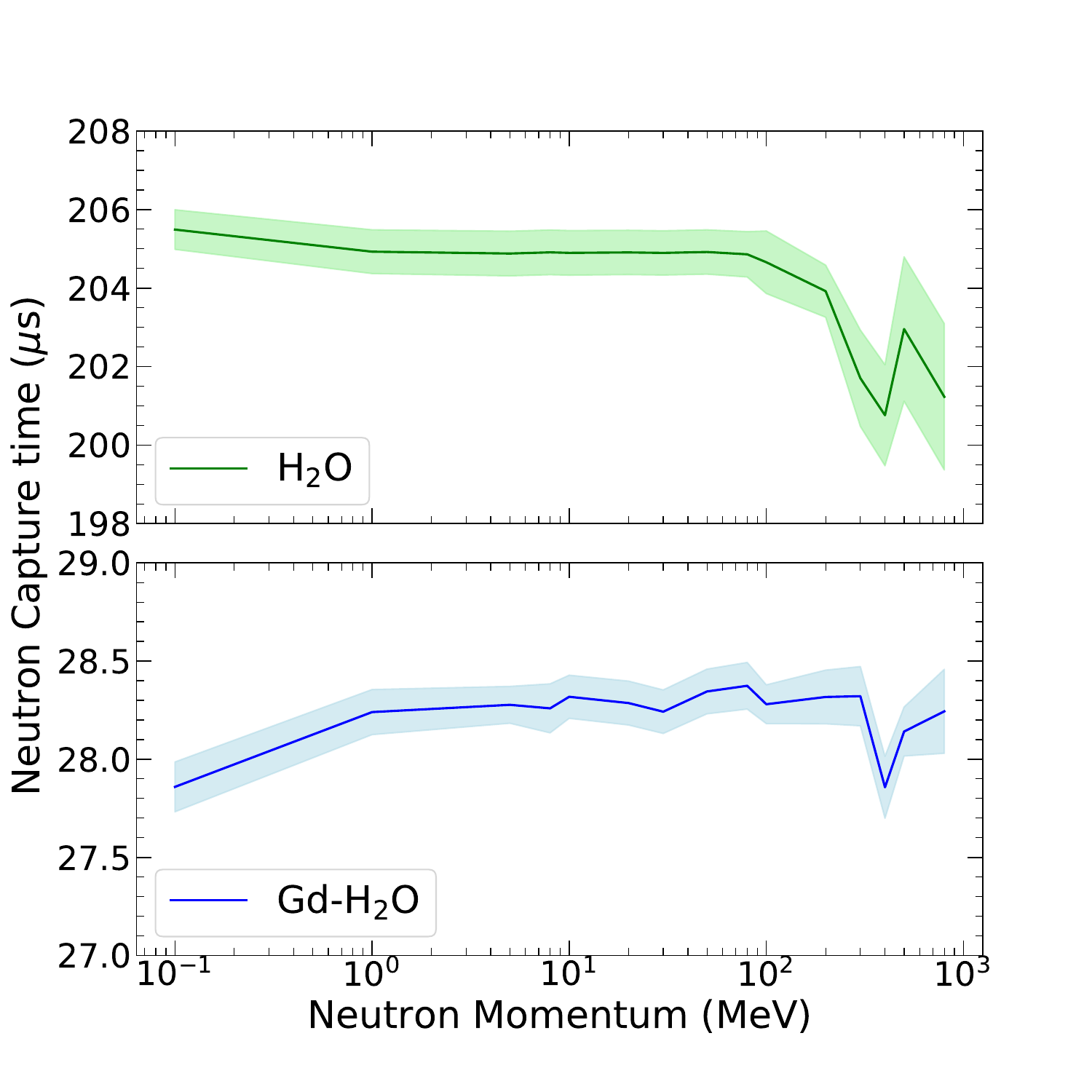}
   \caption{The neutron capture time as a  function of  neutron momentum with error bands representing the standard deviations in both water~(top) and Gd-loaded water~(bottom).}
    \label{fig:gwt}
\end{figure}

Figure~\ref{fig:gwt} shows the neutron capture time as a function of neutron momentum in both water and Gd-loaded water.  For each momentum considered, we simulate $10^5$ neutron events, and select single neutron capture events. 
The neutron capture time between neutron emission and capture is approximately 200 $\mu$s in pure water and 30 $\mu$s in Gd-loaded water, respectively, aligning closely with the findings in the references~\cite{Cokinos:1977zz, Super-Kamiokande:2008mmn, Renshaw:2012np}.  We find that the capture time does not change much with respect to neutron momentum and is therefore unlikely to be useful for event reconstruction.

\subsection{Neutron Capture Distance}

\begin{figure}[t]
    \centering
    \includegraphics[width=3.5in,height=2.9in]{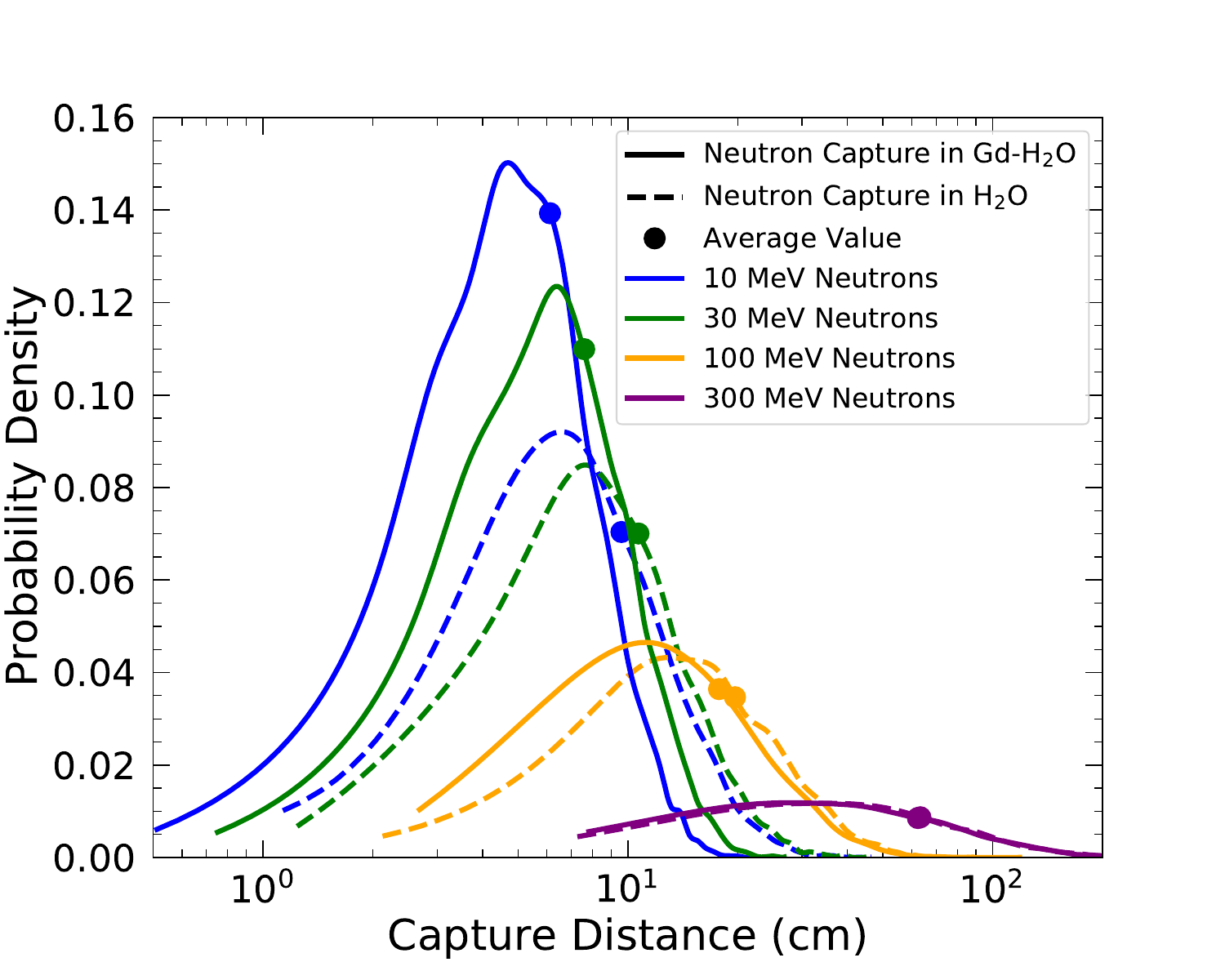}
    \caption{The capture distance of neutrons with different momenta in both water~(dashed lines) and Gd-loaded water~(solid lines). The dot points represent the corresponding average values.}
    \label{fig:capw}
\end{figure}

\begin{figure}[t]
    \centering
    \includegraphics[width=3.5in,height=2.9in]{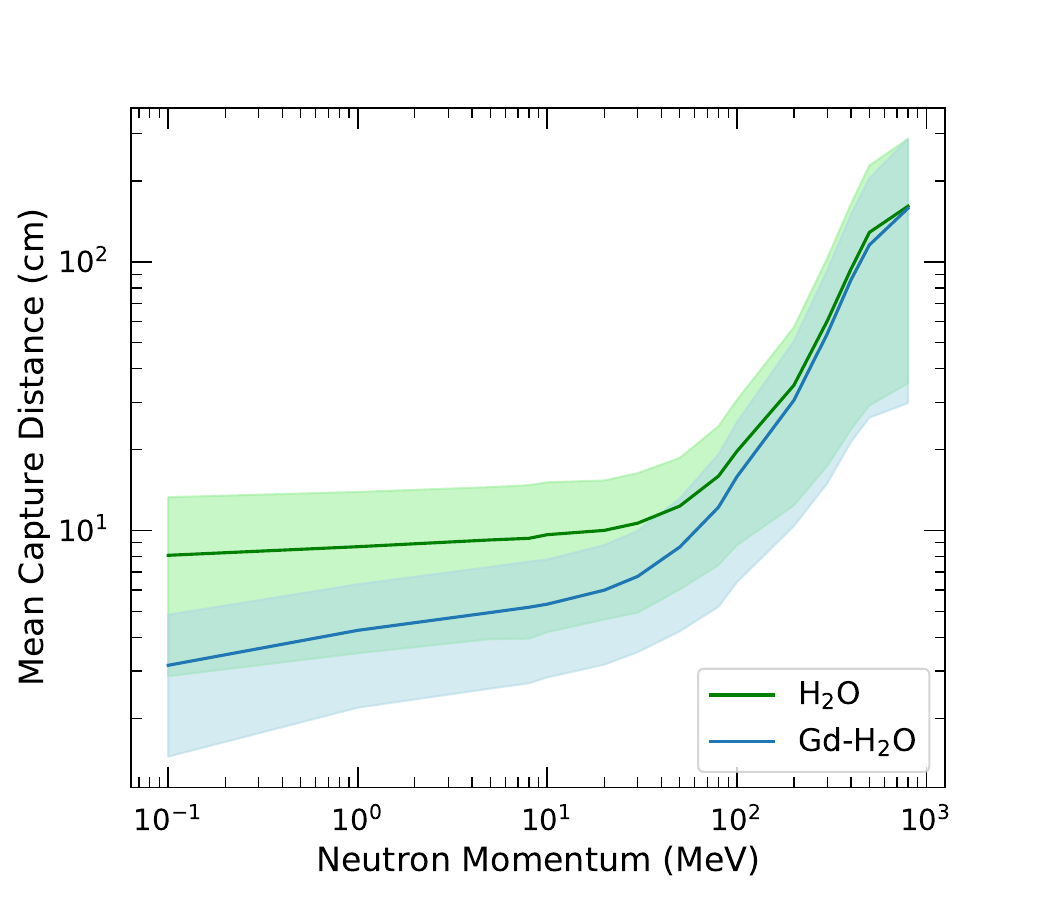}
    \caption{The average neutron capture distances~(solid lines) as  a  function  of  neutron momentum with error bands representing the standard deviations in both water~(green) and Gd-loaded water~(blue).}
    \label{fig:acd}
\end{figure}
The average capture distance for neutrons with a momentum of 30 MeV is approximately 10.6 cm in pure water, which is slightly longer than the average capture distance of 7.5 cm in Gd-loaded water, as shown in Fig.~\ref{fig:gdwd}.

Figure~\ref{fig:capw} shows the distributions of neutron capture distances for several momentum values. There is clearly a correlation of the capture distance with the neutron momentum. 

Figure~\ref{fig:acd} shows the neutron capture distance as a function of neutron momentum in both pure water and Gd-loaded water. 
We find that below about 30\,MeV, the capture distance does not change much with respect to the neutron momentum, but increases rapidly above that. Therefore, the capture distance could potentially be used as a proxy for the neutron momentum.  We do not explore this information further in this work, and defer it to future work. 

\section{Differential Cross Section}
\label{appendix:xsection}
For a given neutrino energy, there is a range of possible positron energies given by the IBD differential cross section~\cite{Strumia:2003zx, Ricciardi:2022pru,Tomalak:2025jtn}, and therefore, a range of possible neutron momenta and angular distributions. 

\begin{figure}[htbp]
    \centering
    \includegraphics[width=3.50in,height=2.9in]{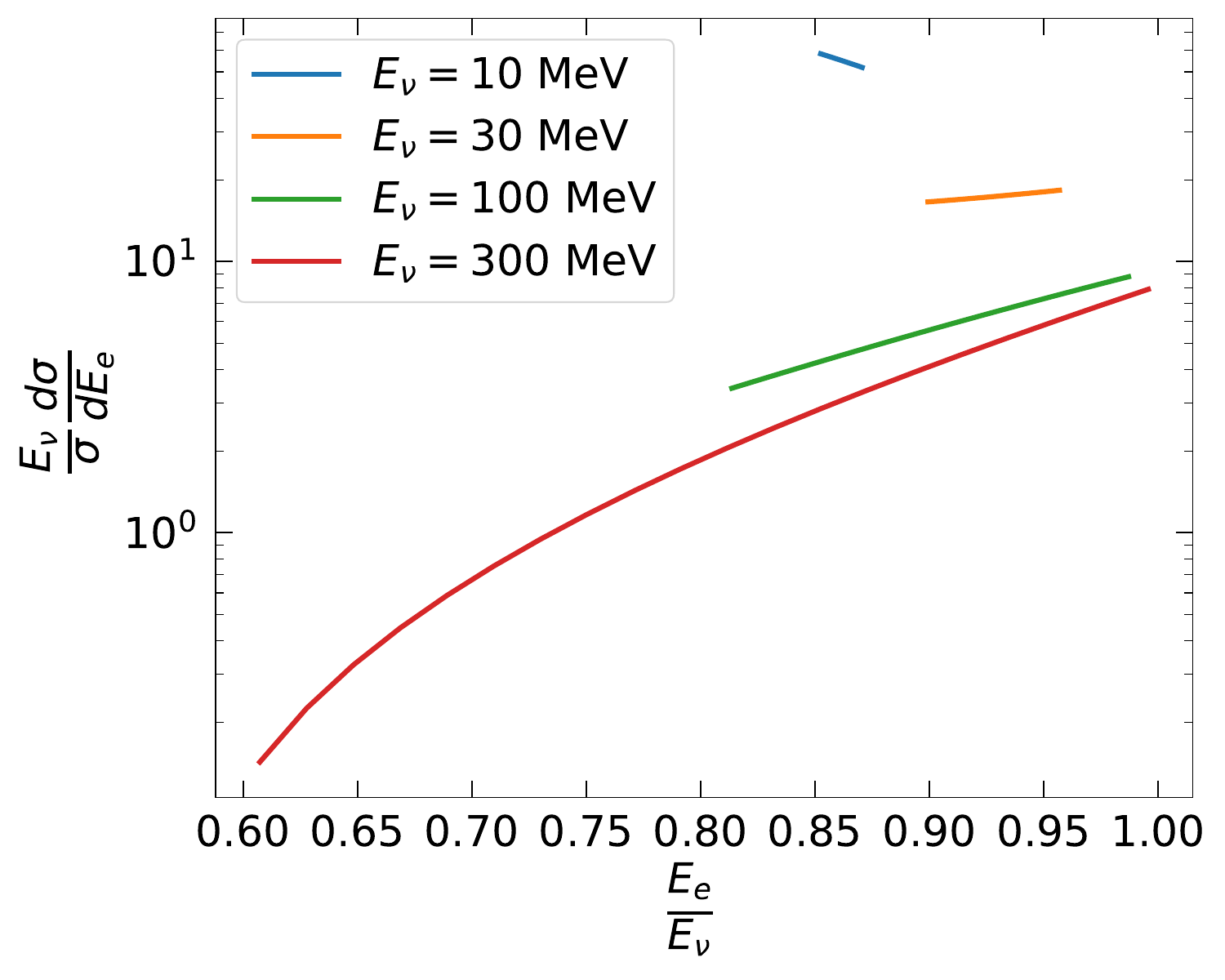}
    \caption{The distribution of positron energies $ \frac{1}{\sigma}\frac{d\sigma}{dE_e}$ generated by cross section for neutrino energies of 10, 30, 100 and 300 MeV. }
    \label{fig:30ee}
\end{figure}

Figure~\ref{fig:30ee} shows the normalized positron energy distribution for several neutrino energies. For higher neutrino energies, the positron energy distribution is broader. 

\begin{figure}[htbp]
    \centering
    \includegraphics[width=3.50in,height=2.9in]{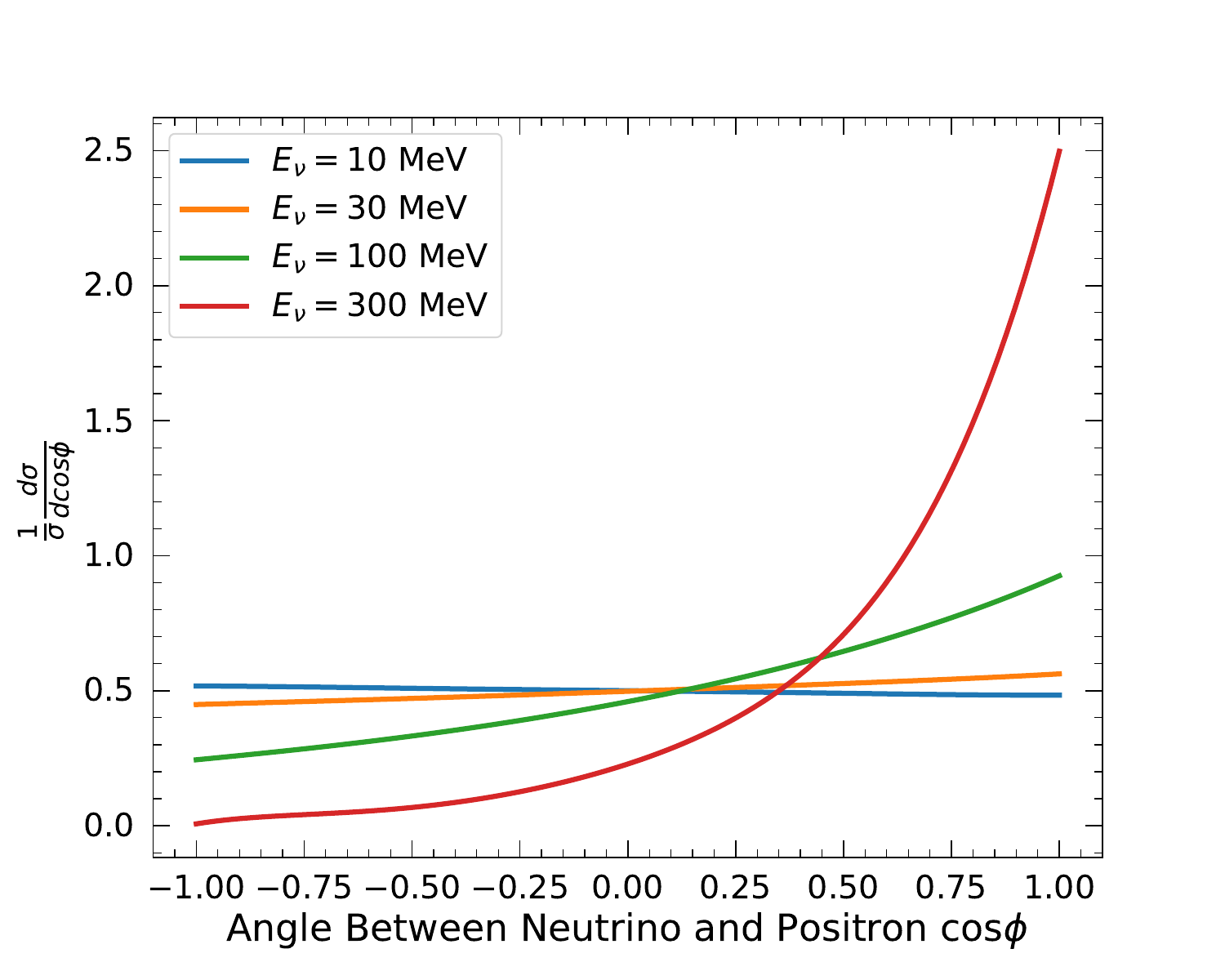}
    \caption{The distribution of positron direction with respect to neutrino direction $ \frac{1}{\sigma}\frac{d\sigma}{d\cos \phi }$, as determined by the cross section, for neutrino energies of 10, 30, 100, and 300 MeV.}
    \label{fig:30p}
\end{figure}

Figure~\ref{fig:30p} shows the angular distribution of positrons relative to the neutrino direction for various neutrino energies. For 10 MeV neutrinos, positrons exhibit a slight backward bias in their emission direction~\cite{Vogel:1999zy}. As the neutrino energy increases, the positron angular distribution shifts towards the forward direction.  As shown here, if only positrons are visible in the IBD, the broad angular distribution of the positrons makes it difficult to meaningfully reconstruct the neutrino direction.  At higher energies, the situation improves. 


\section{Reconstruction Using the
Fixed Positron Energy}
\label{appendix:positron}

In this section, we describe our methodology for reconstructing neutrino directionality, assuming a monochromatic neutrino energy distribution and utilizing the corresponding average positron energy. As an example, starting with a neutrino energy of 30 MeV and the corresponding average positron energy of 27.86 MeV, we derive a neutron momentum of 39.95 MeV. We therefore assign this derived momentum to the nearest available simulation sample, that is, the 40 MeV neutron momentum sample with $10^4$ capture events generated in Geant4. The capture positions of these neutrons are then used to calculate the neutron capture angles $\theta'$ relative to the positron direction for reconstruction.

\begin{figure}[h!]
    \centering
    \includegraphics[width=3.50in,height=3in]{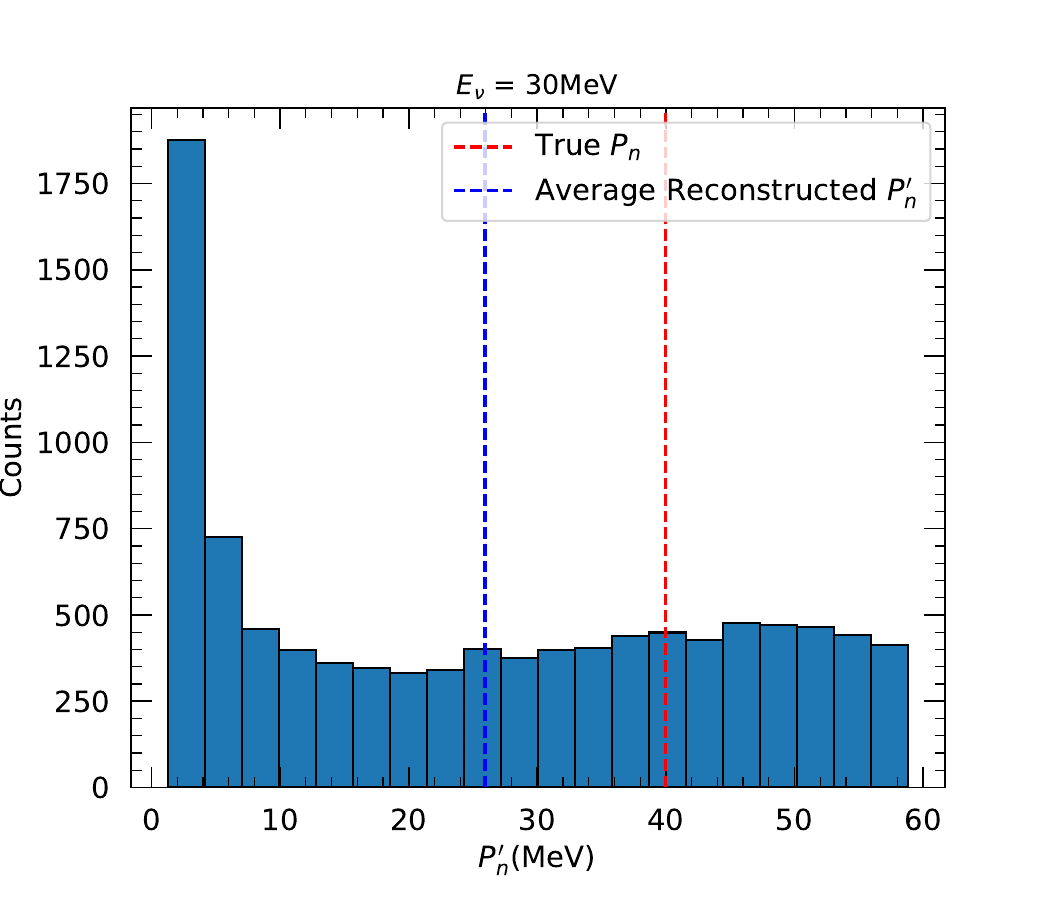}
    \caption{The reconstruction of neutron momentum for a neutrino energy of 30 MeV with the average positron energy. The true neutron momentum and the average reconstructed neutron momentum are shown by the red and blue dashed lines, respectively. }
    \label{fig:30apn}
\end{figure}

\begin{figure}[h!]
    \centering
    \includegraphics[width=3.50in,height=3in]{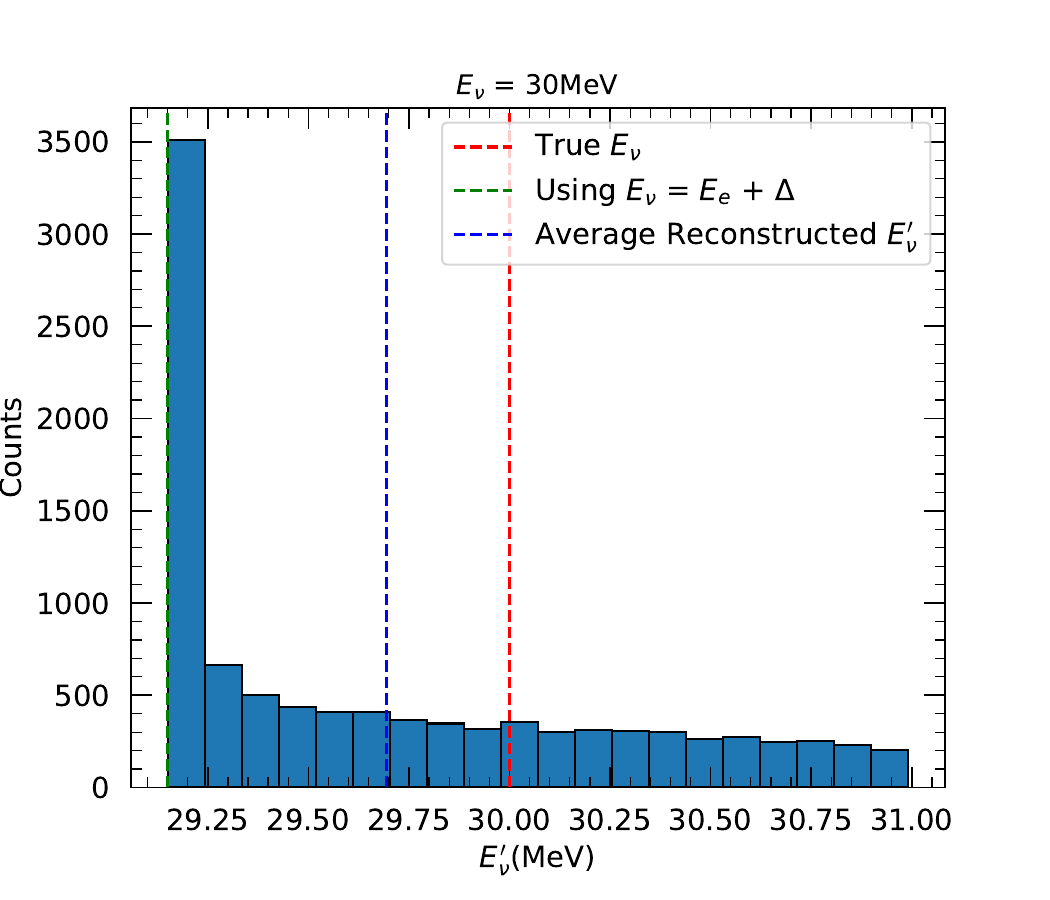}
    \caption{The reconstruction of neutrino energy for a neutrino energy of 30 MeV with the average positron energy. The true neutrino energy, the approximation formula for neutrino energy and the average reconstructed neutrino energy are shown by the red, green and blue dashed lines, respectively. }
    \label{fig:30apv}
\end{figure}

To determine the neutron momentum $P_n'$ and the neutrino energy $E_\nu'$, we substitute the average positron energy $\bar{E_e}$ and the angle $\theta'$ into Eq.~(\ref{eq:ibd}). The reconstructed values of $P_n'$ and $E_\nu'$ are shown in Fig.~\ref{fig:30apn} and Fig.~\ref{fig:30apv}, respectively. The true values and average reconstructed values of neutron momentum and neutrino energy are presented as the red dashed lines. Additionally, for the neutrino energy, we display the approximation formula relating neutrino energy to positron energy~\cite{Vogel:1999zy}, as shown in the green dashed line, 
\begin{equation}
E_\nu=E_e+\Delta,
\end{equation} 
where $\Delta=m_n-m_p \approx 1.293 \mathrm{MeV}$ is the mass difference between neutron and proton.
Directly solving the equations to obtain the neutrino direction is not feasible as it is no longer a 2-dimensional problem. Instead, we ascertain the neutron momentum direction from the capture position and combine it with the magnitude of $P_n'$ to form the neutron momentum vector. By summing this vector with the positron momentum, we can reconstruct the neutrino direction.
Fig.~\ref{fig:30ave} illustrates the reconstruction of the neutrino direction for a 30 MeV neutrino energy. The red dashed line represents the 68\% quantile of the neutrino direction, while the blue dashed line indicates the average reconstructed angle. The 68\% quantile of the reconstructed angle is defined as the value above which 68\% of the data points lie. 

A noticeable peak in Fig.~\ref{fig:30ave} is observed, attributed to the small reconstructed neutron momentum, as indicated by the peak in Fig.~\ref{fig:30apn}, resulting in the dominance of the positron momentum. For different neutrino energies, the reconstruction of neutrino direction resolution using the average positron energy $\cos\psi(\bar{E_e})$ is presented in Table~\ref{tab:diffu}.

\section{Uncertainty from Form Factor}
\label{appendix:form}
The uncertainty in the cross section is dependent on factors such as the CKM matrix element and the form factors~\cite{Strumia:2003zx, Ricciardi:2022pru}. Although the uncertainty in the CKM matrix element can affect the total cross section, it does not influence our reconstruction process. In our analysis, the primary uncertainty in the IBD cross section arises from the form factor $g_1$, which in turn depends on $M_A$~\cite{Strumia:2003zx, Ricciardi:2022pru}. For the parameter $M_A$, we consider a value of $M_A = 1.014 \pm 0.014$ GeV~\cite{Bodek:2007ym}. By utilizing the central value and the associated uncertainty, we can estimate the reconstruction uncertainty arising from variations in the cross section.

\begin{table}[h]
\centering
\begin{tabular}{|c|c|c|}
\hline
$E_\nu$ & Uncertainty of $\cos\psi_{ave}$ & Uncertainty of $\cos\psi_{68\%}$ \\
\hline
10 MeV& 0.0082\% &  0.036\% \\
\hline
30 MeV&  0.019\% & 0.29\%  \\
\hline
100 MeV&  0.043\% &  0.11\% \\
\hline
300 MeV&  0.024\% &  0.070\% \\
\hline
\end{tabular}
\caption{Considering the uncertainty from the cross section, we evaluate the uncertainty in neutrino direction reconstruction across varying neutrino energies.}
    \label{tab:unc}
\end{table}
The analysis results with varying neutrino energies presented in Table~\ref{tab:unc} demonstrate that the reconstruction uncertainty stemming from the cross section is negligible, accounting for less than 1\% of the overall result. Consequently, we exclude this effect in our reconstruction.
\section{Detector Resolution for the Positron}
\label{appendix:positronresolution}
\subsection{Positron Angular Resolution}

The 1$\sigma$ angular resolution refers to the angular range that encompasses 68\% of events in the distribution of the angular difference between the reconstructed direction and the true direction of an event~\cite{Super-Kamiokande:2023jbt}. Positrons with MeV-level kinetic energy typically travel a few centimeters in water before annihilation and we assume that the properties of Cherenkov light that positron and electron produce are identical. The dependence of the angular resolution on electron energy can be found in~\cite{Super-Kamiokande:2010tar,Super-Kamiokande:2023jbt}.

To account for positron angular resolution, we slightly adjust the positron vector (1,0,0) for each neutron capture realization.
We generate a single point $\boldsymbol{a} = (y, z)$ from two normal distributions using eq.~(\ref{eq:normal}) centered at the origin $\boldsymbol{\mu} = (0, 0)$ with a specified standard deviation,  \(\sigma = \frac{\tan(\delta\alpha)}{\chi^2_{0.68, 2}}\), where $\delta \alpha$ is the angular resolution and \(\chi^2_{0.68, 2}\) is the chi-squared value for 68\% probability with 2 degrees of freedom. Finally, we construct a 3-dimensional unit vector \((1, y, z)\) by appending a 1 as the first element and normalizing the resulting vector. This vector represents the measured direction of the positron momentum. This approach incorporates the positron angular resolution by adjusting the positron momentum direction, thereby altering the neutron capture angle $\theta'$ relative to the positron direction.
Table~\ref{tab:vera} displays the 68\% quantile neutrino direction reconstruction resolutions $\cos\psi_{68\%}$ with a 0$^\circ$ and a 10$^\circ$ positron angular resolution at different neutrino energies. The impact of the positron angular resolution is found to be less significant compared to that of vertex resolutions in Table~\ref{tab:verd}.

\begin{table}[h!]
\centering
\begin{tabular}{|c|c|c|c|c|}
\hline
\diagbox{$E_\nu$}{$\cos\psi_{68\%}$}& $\delta \alpha$=0$^\circ$ & $\delta \alpha$=10$^\circ$ \\
\hline
10 MeV& -0.097 & -0.10  \\
\hline
30 MeV& 0.036 & 0.034  \\
\hline

100 MeV & 0.40 & 0.40 \\
\hline

300 MeV& 0.69 & 0.68 \\
\hline
\end{tabular}
\caption{The 68\% quantile neutrino direction reconstruction resolutions for varying neutrino energies, considering positron angular resolution of 10$^\circ$.}  
    \label{tab:vera}
\end{table}

\subsection{Positron Energy Resolution}
In experimental settings, the positron energy resolution $\delta E$ depends on the positron energy~\cite{Super-Kamiokande:2023jbt}. For simplification, we consider a constant positron energy resolution $\delta E$ of 10\%. Similar to above, we modify the positron energy according to the energy resolution for each neutron capture realization. 
We generate a random number $p$ drawn from a normal distribution with a mean of 0 and a standard deviation of 10\%. We then apply this random fluctuation to the true positron energy, such that the reconstructed positron energy is given by $E_e' =E_e \times (1 + p) $. The reconstruction results are presented in Table~\ref{tab:vere}. The impact of the positron energy resolution of 10\% is not significant. 
\begin{table}[h!]
\centering
\begin{tabular}{|c|c|c|c|c|}
\hline
\diagbox{$E_\nu$}{$\cos\psi_{68\%}$}& $\delta E$=0\% & $\delta E$=10\% \\
\hline
10  MeV& -0.097 & -0.098 \\
\hline
30  MeV& 0.036 & 0.036 \\
\hline
100 MeV& 0.40 & 0.40 \\
\hline
300 MeV& 0.69 & 0.69 \\
\hline
\end{tabular}
\caption{The 68\% quantile neutrino direction reconstruction resolutions for varying neutrino energies, considering positron energy resolution of 10 \%.}
    \label{tab:vere}
\end{table}

\subsection{Reconstruction Using the Fixed Positron Energy with Detector Resolutions}
\begin{figure}[htbp]
    \centering
    \includegraphics[width=3.50in,height=3in]{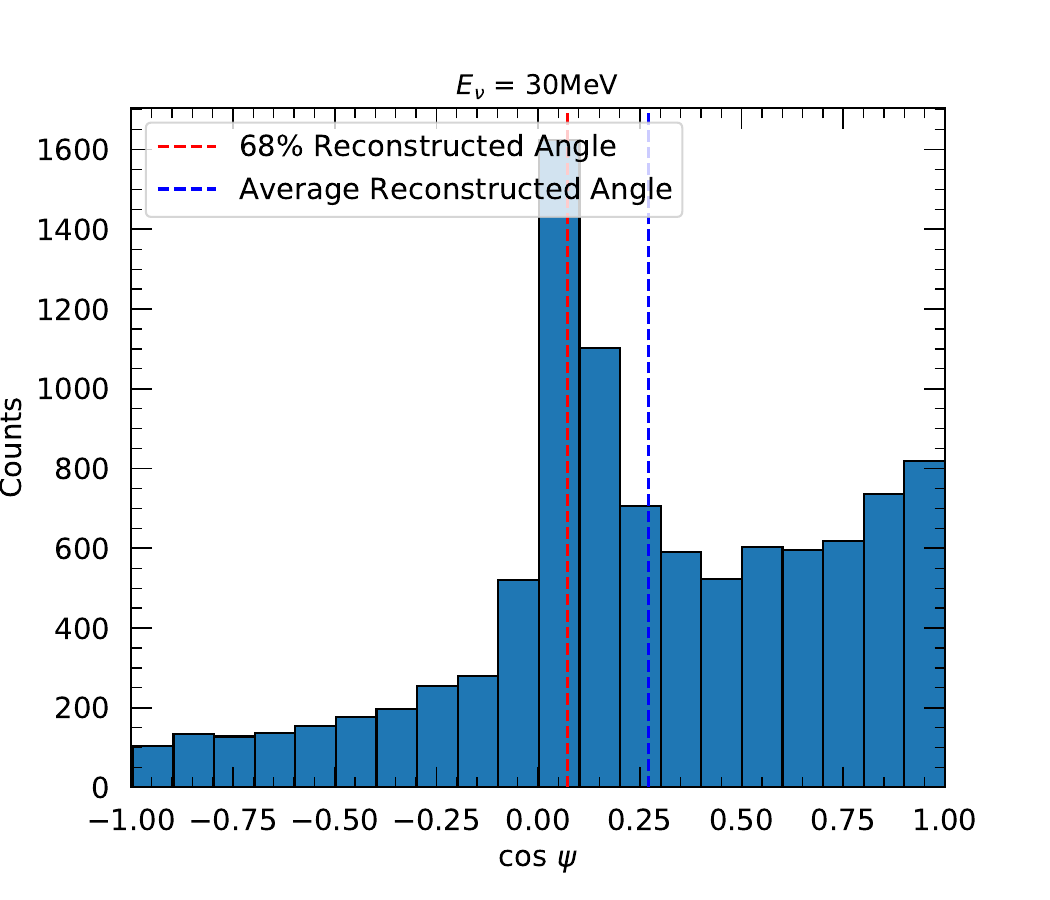}
    \caption{The reconstruction of neutrino direction resolution for a neutrino energy of 30 MeV with corresponding average positron energy. The 68\% quantile of neutrino direction resolution and the average reconstructed resolution angle are shown by the red and blue dashed line, respectively. }
    \label{fig:30ave}
\end{figure}

\begin{figure}[htbp]
    \centering
    \includegraphics[width=3.50in,height=3in]{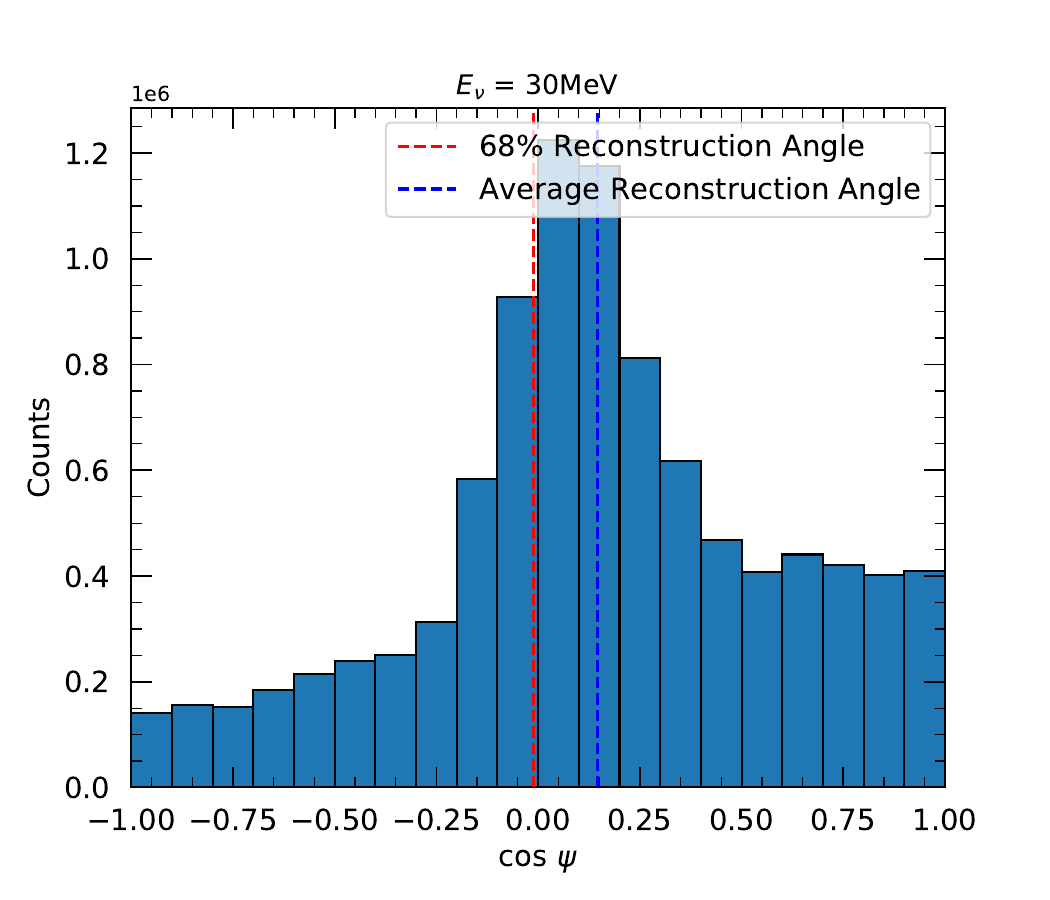}
    \caption{ The reconstruction of neutrino direction resolution for a neutrino energy of 30 MeV with corresponding average positron energy, where the neutron capture and inverse beta decay vertex resolutions of 10 cm, positron angular resolution of 10$^\circ$, and positron energy resolution of 10\% are applied. The 68\% quantile of neutrino direction resolution and the average reconstructed resolution angle are shown by the red and blue dashed line, respectively. }
    \label{fig:30aver}
\end{figure}

Consider a scenario where a neutron capture and IBD vertex resolution of 10 cm, a positron angular resolution of 10$^\circ$, and a positron energy resolution of 10\% are incorporated. Fig.~\ref{fig:30aver} illustrates the reconstructed neutrino direction and corresponding mean positron energy for 30 MeV neutrinos under these conditions. Compared to the idealized case shown in Fig.~\ref{fig:30ave}, there are more counts at smaller values of $\cos \psi$ and the average value and 68\% quantile of neutrino direction resolution are also smaller, indicating a deterioration in angular reconstruction after considering these factors, as expected.
\begin{table}[h]
\centering
\begin{tabular}{|c|c|c|c|c|}
\hline
\diagbox{$E_\nu$}{$\cos\psi$} & $\cos\psi_{ave}$ $(\bar{E_e})$ & $\cos\psi_{68\%}$ $(\bar{E_e})$ & $\cos\psi_{ave}$ & $\cos\psi_{68\%}$ \\
\hline
10 MeV& 0.17 & -0.014 & 0.19 & -0.097 \\
\hline
30 MeV& 0.27 & 0.073 & 0.28 & 0.036 \\
\hline
100 MeV& 0.48 & 0.28 & 0.50 & 0.40 \\
\hline
300 MeV& 0.65 & 0.64 & 0.69 & 0.69 \\
\hline
\end{tabular}
\caption{The reconstruction of neutrino direction for varying neutrino energy by using the corresponding average positron energy $\cos\psi(\bar{E_e})$ and incorporating information from the cross section.}
    \label{tab:diffu}
\end{table}

\section{The Average Value of Neutrino Direction Resolution}
\label{appendix:resolution}

\begin{figure*}[htbp]
  \centering
  
    \includegraphics[width=\textwidth]{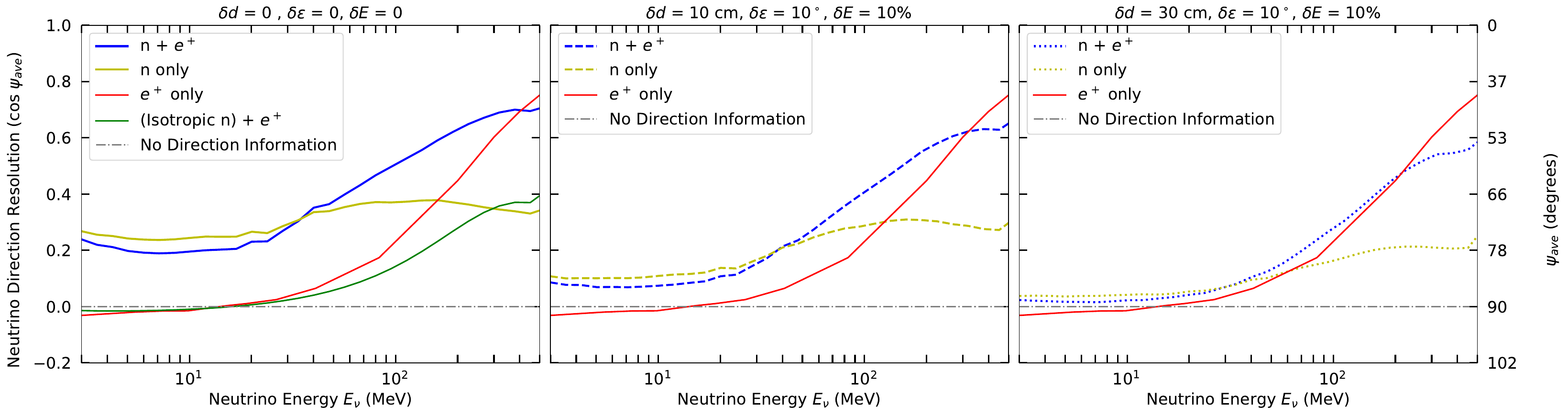}

  \caption{The average value of the neutrino direction reconstruction resolution is shown as a function of neutrino energy from 3 MeV to 500 MeV, for different resolution scenarios. In the left plot, the blue solid curve shows the reconstruction using positron and neutron diffusion information with perfect detector resolution. The yellow curve represents experiments considering only the neutron capture position, excluding positron data. The red solid curve depicts the performance using only the positron direction, without accounting for any detector resolutions. The green curve follows the same method as the blue curve but assumes neutrons are emitted isotropically. The gray dot-dashed line indicates the average value for the scenario without any directional information. The middle and right plots incorporate additional detector resolutions: a neutron capture and IBD vertex resolution of 10 cm and 30 cm, respectively, a positron angular resolution of 10$^\circ$, and a positron energy resolution of 10\%.}

  \label{fig:300reca}
\end{figure*}

In addition to the 68\% quantile of the neutrino direction resolution shown in the main text, we also provide the average neutrino direction resolution as a function of neutrino energy from 3 MeV to 500 MeV for different scenarios, as shown in Fig.~\ref{fig:300reca}.
In regions of high neutrino energy, the positron direction resolution in Fig.~\ref{fig:300reca} is better than that of this method. This is because diffusion causes some directional changes, and during event selection, only events with one neutron capture are chosen. As a result, lower-energy neutrons are given more weight in the reconstruction, leading to some bias.
\acknowledgments
We thank Shirley Li for coming up with this idea together many years ago. 
We are grateful for the helpful discussions with CHU Ming Chung, Bei Zhou, John F. Beacom, YIP Chun Ming Ivan, LEONG Hin Wai, Tse Shing Him Samuel, Weizhen Zhang, Fong Chingam and Yufeng Li. The works of QSL and KCYN are supported by Croucher foundation, RGC grants (24302721, 14305822, 14308023), Joint NSFC/GRC grant (N CUHK456/22), and NSFC grant 12322517. This work is supported in part by the National Natural Science Foundation of China under grant No. 12342502.
\bigskip

\bibliographystyle{jhep}
\bibliography{reference/reference}

\providecommand{\href}[2]{#2}\begingroup\raggedright\begin{thebibliography}{10}

\bibitem{Bahcall:1987jc}
J.~N. Bahcall and R.~K. Ulrich, \emph{{Solar Models, Neutrino Experiments and Helioseismology}}, \href{https://doi.org/10.1103/RevModPhys.60.297}{\emph{Rev. Mod. Phys.} {\bfseries 60} (1988) 297}.

\bibitem{Kamiokande-II:1987idp}
{\scshape Kamiokande-II} collaboration, \emph{{Observation of a Neutrino Burst from the Supernova SN 1987a}}, \href{https://doi.org/10.1103/PhysRevLett.58.1490}{\emph{Phys. Rev. Lett.} {\bfseries 58} (1987) 1490}.

\bibitem{Hirata:1988ad}
K.~S. Hirata et~al., \emph{{Observation in the Kamiokande-II Detector of the Neutrino Burst from Supernova SN 1987a}}, \href{https://doi.org/10.1103/PhysRevD.38.448}{\emph{Phys. Rev. D} {\bfseries 38} (1988) 448}.

\bibitem{Bionta:1987qt}
R.~M. Bionta et~al., \emph{{Observation of a Neutrino Burst in Coincidence with Supernova SN 1987a in the Large Magellanic Cloud}}, \href{https://doi.org/10.1103/PhysRevLett.58.1494}{\emph{Phys. Rev. Lett.} {\bfseries 58} (1987) 1494}.

\bibitem{IMB:1988suc}
{\scshape IMB} collaboration, \emph{{Angular Distribution of Events From Sn1987a}}, \href{https://doi.org/10.1103/PhysRevD.37.3361}{\emph{Phys. Rev. D} {\bfseries 37} (1988) 3361}.

\bibitem{Alekseev:1988gp}
E.~N. Alekseev, L.~N. Alekseeva, I.~V. Krivosheina and V.~I. Volchenko, \emph{{Detection of the Neutrino Signal From {SN1987A} in the {LMC} Using the Inr Baksan Underground Scintillation Telescope}}, \href{https://doi.org/10.1016/0370-2693(88)91651-6}{\emph{Phys. Lett. B} {\bfseries 205} (1988) 209}.

\bibitem{Qian:2018wid}
X.~Qian and J.-C. Peng, \emph{{Physics with Reactor Neutrinos}}, \href{https://doi.org/10.1088/1361-6633/aae881}{\emph{Rept. Prog. Phys.} {\bfseries 82} (2019) 036201} [\href{https://arxiv.org/abs/1801.05386}{{\ttfamily 1801.05386}}].

\bibitem{Fiorentini:2007te}
G.~Fiorentini, M.~Lissia and F.~Mantovani, \emph{{Geo-neutrinos and Earth's interior}}, \href{https://doi.org/10.1016/j.physrep.2007.09.001}{\emph{Phys. Rept.} {\bfseries 453} (2007) 117} [\href{https://arxiv.org/abs/0707.3203}{{\ttfamily 0707.3203}}].

\bibitem{Super-Kamiokande:1998kpq}
{\scshape Super-Kamiokande} collaboration, \emph{{Evidence for oscillation of atmospheric neutrinos}}, \href{https://doi.org/10.1103/PhysRevLett.81.1562}{\emph{Phys. Rev. Lett.} {\bfseries 81} (1998) 1562} [\href{https://arxiv.org/abs/hep-ex/9807003}{{\ttfamily hep-ex/9807003}}].

\bibitem{Kajita:2010zz}
T.~Kajita, \emph{{ATMOSPHERIC NEUTRINOS AND DISCOVERY OF NEUTRINO OSCILLATIONS}}, \href{https://doi.org/10.2183/pjab.86.303}{\emph{Proc. Japan Acad. B} {\bfseries 86} (2010) 303}.

\bibitem{Zhou:2023mou}
B.~Zhou and J.~F. Beacom, \emph{{First detailed calculation of atmospheric neutrino foregrounds to the diffuse supernova neutrino background in Super-Kamiokande}}, \href{https://doi.org/10.1103/PhysRevD.109.103003}{\emph{Phys. Rev. D} {\bfseries 109} (2024) 103003} [\href{https://arxiv.org/abs/2311.05675}{{\ttfamily 2311.05675}}].

\bibitem{Muhlbeier:2013gwa}
T.~M\"uhlbeier, H.~Nunokawa and R.~Zukanovich~Funchal, \emph{{Revisiting the Triangulation Method for Pointing to Supernova and Failed Supernova with Neutrinos}}, \href{https://doi.org/10.1103/PhysRevD.88.085010}{\emph{Phys. Rev. D} {\bfseries 88} (2013) 085010} [\href{https://arxiv.org/abs/1304.5006}{{\ttfamily 1304.5006}}].

\bibitem{Brdar:2018zds}
V.~Brdar, M.~Lindner and X.-J. Xu, \emph{{Neutrino astronomy with supernova neutrinos}}, \href{https://doi.org/10.1088/1475-7516/2018/04/025}{\emph{JCAP} {\bfseries 04} (2018) 025} [\href{https://arxiv.org/abs/1802.02577}{{\ttfamily 1802.02577}}].

\bibitem{Hansen:2019giq}
R.~S.~L. Hansen, M.~Lindner and O.~Scholer, \emph{{Timing the neutrino signal of a Galactic supernova}}, \href{https://doi.org/10.1103/PhysRevD.101.123018}{\emph{Phys. Rev. D} {\bfseries 101} (2020) 123018} [\href{https://arxiv.org/abs/1904.11461}{{\ttfamily 1904.11461}}].

\bibitem{Beacom:1998fj}
J.~F. Beacom and P.~Vogel, \emph{{Can a supernova be located by its neutrinos?}}, \href{https://doi.org/10.1103/PhysRevD.60.033007}{\emph{Phys. Rev. D} {\bfseries 60} (1999) 033007} [\href{https://arxiv.org/abs/astro-ph/9811350}{{\ttfamily astro-ph/9811350}}].

\bibitem{Laha:2013hva}
R.~Laha and J.~F. Beacom, \emph{{Gadolinium in water Cherenkov detectors improves detection of supernova $\nu_e$}}, \href{https://doi.org/10.1103/PhysRevD.89.063007}{\emph{Phys. Rev. D} {\bfseries 89} (2014) 063007} [\href{https://arxiv.org/abs/1311.6407}{{\ttfamily 1311.6407}}].

\bibitem{Adams:2013ana}
S.~M. Adams, C.~S. Kochanek, J.~F. Beacom, M.~R. Vagins and K.~Z. Stanek, \emph{{Observing the Next Galactic Supernova}}, \href{https://doi.org/10.1088/0004-637X/778/2/164}{\emph{Astrophys. J.} {\bfseries 778} (2013) 164} [\href{https://arxiv.org/abs/1306.0559}{{\ttfamily 1306.0559}}].

\bibitem{Formaggio:2012cpf}
J.~A. Formaggio and G.~P. Zeller, \emph{{From eV to EeV: Neutrino Cross Sections Across Energy Scales}}, \href{https://doi.org/10.1103/RevModPhys.84.1307}{\emph{Rev. Mod. Phys.} {\bfseries 84} (2012) 1307} [\href{https://arxiv.org/abs/1305.7513}{{\ttfamily 1305.7513}}].

\bibitem{Ricciardi:2022pru}
G.~Ricciardi, N.~Vignaroli and F.~Vissani, \emph{{An accurate evaluation of electron (anti-)neutrino scattering on nucleons}}, \href{https://doi.org/10.1007/JHEP08(2022)212}{\emph{JHEP} {\bfseries 08} (2022) 212} [\href{https://arxiv.org/abs/2206.05567}{{\ttfamily 2206.05567}}].

\bibitem{Mirizzi:2015eza}
A.~Mirizzi, I.~Tamborra, H.-T. Janka, N.~Saviano, K.~Scholberg, R.~Bollig et~al., \emph{{Supernova Neutrinos: Production, Oscillations and Detection}}, \href{https://doi.org/10.1393/ncr/i2016-10120-8}{\emph{Riv. Nuovo Cim.} {\bfseries 39} (2016) 1} [\href{https://arxiv.org/abs/1508.00785}{{\ttfamily 1508.00785}}].

\bibitem{Araki:2005qa}
T.~Araki et~al., \emph{{Experimental investigation of geologically produced antineutrinos with KamLAND}}, \href{https://doi.org/10.1038/nature03980}{\emph{Nature} {\bfseries 436} (2005) 499}.

\bibitem{JUNO:2025gmd}
{\scshape JUNO} collaboration, \emph{{First measurement of reactor neutrino oscillations at JUNO}},  \href{https://arxiv.org/abs/2511.14593}{{\ttfamily 2511.14593}}.

\bibitem{JUNO:2015zny}
{\scshape JUNO} collaboration, \emph{{Neutrino Physics with JUNO}}, \href{https://doi.org/10.1088/0954-3899/43/3/030401}{\emph{J. Phys. G} {\bfseries 43} (2016) 030401} [\href{https://arxiv.org/abs/1507.05613}{{\ttfamily 1507.05613}}].

\bibitem{Yuksel:2007ac}
H.~Yuksel, S.~Horiuchi, J.~F. Beacom and S.~Ando, \emph{{Neutrino Constraints on the Dark Matter Total Annihilation Cross Section}}, \href{https://doi.org/10.1103/PhysRevD.76.123506}{\emph{Phys. Rev. D} {\bfseries 76} (2007) 123506} [\href{https://arxiv.org/abs/0707.0196}{{\ttfamily 0707.0196}}].

\bibitem{Palomares-Ruiz:2007trf}
S.~Palomares-Ruiz and S.~Pascoli, \emph{{Testing MeV dark matter with neutrino detectors}}, \href{https://doi.org/10.1103/PhysRevD.77.025025}{\emph{Phys. Rev. D} {\bfseries 77} (2008) 025025} [\href{https://arxiv.org/abs/0710.5420}{{\ttfamily 0710.5420}}].

\bibitem{Liu:2023cqs}
Q.~Liu and K.~C.~Y. Ng, \emph{{Sensitivity floor for primordial black holes in neutrino searches}}, \href{https://doi.org/10.1103/PhysRevD.110.063024}{\emph{Phys. Rev. D} {\bfseries 110} (2024) 063024} [\href{https://arxiv.org/abs/2312.06108}{{\ttfamily 2312.06108}}].

\bibitem{Cappiello:2019qsw}
C.~V. Cappiello and J.~F. Beacom, \emph{{Strong New Limits on Light Dark Matter from Neutrino Experiments}}, \href{https://doi.org/10.1103/PhysRevD.104.069901}{\emph{Phys. Rev. D} {\bfseries 100} (2019) 103011} [\href{https://arxiv.org/abs/1906.11283}{{\ttfamily 1906.11283}}].

\bibitem{Arguelles:2019ouk}
C.~A. Arg\"uelles, A.~Diaz, A.~Kheirandish, A.~Olivares-Del-Campo, I.~Safa and A.~C. Vincent, \emph{{Dark matter annihilation to neutrinos}}, \href{https://doi.org/10.1103/RevModPhys.93.035007}{\emph{Rev. Mod. Phys.} {\bfseries 93} (2021) 035007} [\href{https://arxiv.org/abs/1912.09486}{{\ttfamily 1912.09486}}].

\bibitem{Tse:2026ppi}
S.~S.~H. Tse, Q.~Liu, K.~C.~Y. Ng, K.~Choi and Y.-F. Li, \emph{{Boosted Dark Matter Directionality in Large Liquid Scintillator Detectors}},  \href{https://arxiv.org/abs/2610.31666}{{\ttfamily 2610.31666}}.

\bibitem{Vogel:1999zy}
P.~Vogel and J.~F. Beacom, \emph{{Angular distribution of neutron inverse beta decay, anti-neutrino(e) + p ---\ensuremath{>} e+ + n}}, \href{https://doi.org/10.1103/PhysRevD.60.053003}{\emph{Phys. Rev. D} {\bfseries 60} (1999) 053003} [\href{https://arxiv.org/abs/hep-ph/9903554}{{\ttfamily hep-ph/9903554}}].

\bibitem{CHOOZ:1999hgz}
{\scshape CHOOZ} collaboration, \emph{{Determination of neutrino incoming direction in the CHOOZ experiment and supernova explosion location by scintillator detectors}}, \href{https://doi.org/10.1103/PhysRevD.61.012001}{\emph{Phys. Rev. D} {\bfseries 61} (2000) 012001} [\href{https://arxiv.org/abs/hep-ex/9906011}{{\ttfamily hep-ex/9906011}}].

\bibitem{Fischer:2015oma}
V.~Fischer et~al., \emph{{Prompt directional detection of galactic supernova by combining large liquid scintillator neutrino detectors}}, \href{https://doi.org/10.1088/1475-7516/2015/08/032}{\emph{JCAP} {\bfseries 08} (2015) 032} [\href{https://arxiv.org/abs/1504.05466}{{\ttfamily 1504.05466}}].

\bibitem{Mukhopadhyay:2020ubs}
M.~Mukhopadhyay, C.~Lunardini, F.~X. Timmes and K.~Zuber, \emph{{Presupernova neutrinos: directional sensitivity and prospects for progenitor identification}}, \href{https://doi.org/10.3847/1538-4357/ab99a6}{\emph{Astrophys. J.} {\bfseries 899} (2020) 153} [\href{https://arxiv.org/abs/2004.02045}{{\ttfamily 2004.02045}}].

\bibitem{Li:2020gaz}
H.-L. Li, Y.-F. Li, L.-J. Wen and S.~Zhou, \emph{{Prospects for Pre-supernova Neutrino Observation in Future Large Liquid-scintillator Detectors}}, \href{https://doi.org/10.1088/1475-7516/2020/05/049}{\emph{JCAP} {\bfseries 05} (2020) 049} [\href{https://arxiv.org/abs/2003.03982}{{\ttfamily 2003.03982}}].

\bibitem{Beacom:2003nk}
J.~F. Beacom and M.~R. Vagins, \emph{{GADZOOKS! Anti-neutrino spectroscopy with large water Cherenkov detectors}}, \href{https://doi.org/10.1103/PhysRevLett.93.171101}{\emph{Phys. Rev. Lett.} {\bfseries 93} (2004) 171101} [\href{https://arxiv.org/abs/hep-ph/0309300}{{\ttfamily hep-ph/0309300}}].

\bibitem{Suzuki:2019jby}
Y.~Suzuki, \emph{{The Super-Kamiokande experiment}}, \href{https://doi.org/10.1140/epjc/s10052-019-6796-2}{\emph{Eur. Phys. J. C} {\bfseries 79} (2019) 298}.

\bibitem{Takeuchi:2022dfj}
{\scshape Super-Kamiokande} collaboration, \emph{{Recent oscillation results and future prospects of Super-Kamiokande}}, \href{https://doi.org/10.22323/1.421.0004}{\emph{PoS} {\bfseries NOW2022} (2023) 004}.

\bibitem{Super-Kamiokande:2021the}
{\scshape Super-Kamiokande} collaboration, \emph{{First gadolinium loading to Super-Kamiokande}}, \href{https://doi.org/10.1016/j.nima.2021.166248}{\emph{Nucl. Instrum. Meth. A} {\bfseries 1027} (2022) 166248} [\href{https://arxiv.org/abs/2109.00360}{{\ttfamily 2109.00360}}].

\bibitem{Abe:2024ydm}
K.~Abe et~al., \emph{{Second gadolinium loading to Super-Kamiokande}},  \href{https://arxiv.org/abs/2403.07796}{{\ttfamily 2403.07796}}.

\bibitem{Koshio:2025fjs}
Y.~Koshio, M.~Nakahata, H.~Sekiya and M.~R. Vagins, \emph{{Upgrade of Super-Kamiokande with Gadolinium}},  \href{https://arxiv.org/abs/2511.03921}{{\ttfamily 2511.03921}}.

\bibitem{Super-Kamiokande:2002weg}
{\scshape Super-Kamiokande} collaboration, \emph{{The Super-Kamiokande detector}}, \href{https://doi.org/10.1016/S0168-9002(03)00425-X}{\emph{Nucl. Instrum. Meth. A} {\bfseries 501} (2003) 418}.

\bibitem{Li:2014sea}
S.~W. Li and J.~F. Beacom, \emph{{First calculation of cosmic-ray muon spallation backgrounds for MeV astrophysical neutrino signals in Super-Kamiokande}}, \href{https://doi.org/10.1103/PhysRevC.89.045801}{\emph{Phys. Rev. C} {\bfseries 89} (2014) 045801} [\href{https://arxiv.org/abs/1402.4687}{{\ttfamily 1402.4687}}].

\bibitem{Nikrant:2017nya}
A.~Nikrant, R.~Laha and S.~Horiuchi, \emph{{Robust measurement of supernova $\nu_e$ spectra with future neutrino detectors}}, \href{https://doi.org/10.1103/PhysRevD.97.023019}{\emph{Phys. Rev. D} {\bfseries 97} (2018) 023019} [\href{https://arxiv.org/abs/1711.00008}{{\ttfamily 1711.00008}}].

\bibitem{Nairat:2024upg}
O.~Nairat, J.~F. Beacom and S.~W. Li, \emph{{Neutron tagging can greatly reduce spallation backgrounds in Super-Kamiokande}}, \href{https://doi.org/10.1103/PhysRevD.111.023014}{\emph{Phys. Rev. D} {\bfseries 111} (2025) 023014} [\href{https://arxiv.org/abs/2409.10611}{{\ttfamily 2409.10611}}].

\bibitem{Super-Kamiokande:2023xup}
{\scshape Super-Kamiokande} collaboration, \emph{{Search for Astrophysical Electron Antineutrinos in Super-Kamiokande with 0.01\% Gadolinium-loaded Water}}, \href{https://doi.org/10.3847/2041-8213/acdc9e}{\emph{Astrophys. J. Lett.} {\bfseries 951} (2023) L27} [\href{https://arxiv.org/abs/2305.05135}{{\ttfamily 2305.05135}}].

\bibitem{Nakahata:2015vma}
M.~Nakahata, \emph{{Future Prospects of Super-kamiokande and Hyper-kamiokande}}, \href{https://doi.org/10.1016/j.phpro.2014.12.054}{\emph{Phys. Procedia} {\bfseries 61} (2015) 568}.

\bibitem{Renshaw:2012np}
{\scshape Super-Kamiokande} collaboration, \emph{{Research and Development for a Gadolinium Doped Water Cherenkov Detector}}, \href{https://doi.org/10.1016/j.phpro.2012.02.467}{\emph{Phys. Procedia} {\bfseries 37} (2012) 1249} [\href{https://arxiv.org/abs/1201.1017}{{\ttfamily 1201.1017}}].

\bibitem{Super-Kamiokande:2008mmn}
{\scshape Super-Kamiokande} collaboration, \emph{{First Study of Neutron Tagging with a Water Cherenkov Detector}}, \href{https://doi.org/10.1016/j.astropartphys.2009.03.002}{\emph{Astropart. Phys.} {\bfseries 31} (2009) 320} [\href{https://arxiv.org/abs/0811.0735}{{\ttfamily 0811.0735}}].

\bibitem{Allison:2016lfl}
J.~Allison et~al., \emph{{Recent developments in Geant4}}, \href{https://doi.org/10.1016/j.nima.2016.06.125}{\emph{Nucl. Instrum. Meth. A} {\bfseries 835} (2016) 186}.

\bibitem{GEANT4:2002zbu}
{\scshape GEANT4} collaboration, \emph{{GEANT4--a simulation toolkit}}, \href{https://doi.org/10.1016/S0168-9002(03)01368-8}{\emph{Nucl. Instrum. Meth. A} {\bfseries 506} (2003) 250}.

\bibitem{Hino:2024mbb}
Y.~Hino et~al., \emph{{Modification on thermal motion in Geant4 for neutron capture simulation in Gadolinium loaded water}},  \href{https://arxiv.org/abs/2412.04186}{{\ttfamily 2412.04186}}.

\bibitem{Super-Kamiokande:2008ecj}
{\scshape Super-Kamiokande} collaboration, \emph{{Solar neutrino measurements in Super-Kamiokande-II}}, \href{https://doi.org/10.1103/PhysRevD.78.032002}{\emph{Phys. Rev. D} {\bfseries 78} (2008) 032002} [\href{https://arxiv.org/abs/0803.4312}{{\ttfamily 0803.4312}}].

\bibitem{Super-Kamiokande:2010tar}
{\scshape Super-Kamiokande} collaboration, \emph{{Solar neutrino results in Super-Kamiokande-III}}, \href{https://doi.org/10.1103/PhysRevD.83.052010}{\emph{Phys. Rev. D} {\bfseries 83} (2011) 052010} [\href{https://arxiv.org/abs/1010.0118}{{\ttfamily 1010.0118}}].

\bibitem{Super-Kamiokande:2023jbt}
{\scshape Super-Kamiokande} collaboration, \emph{{Solar neutrino measurements using the full data period of Super-Kamiokande-IV}}, \href{https://doi.org/10.1103/PhysRevD.109.092001}{\emph{Phys. Rev. D} {\bfseries 109} (2024) 092001} [\href{https://arxiv.org/abs/2312.12907}{{\ttfamily 2312.12907}}].

\bibitem{Super-Kamiokande:2016yck}
{\scshape Super-Kamiokande} collaboration, \emph{{Solar Neutrino Measurements in Super-Kamiokande-IV}}, \href{https://doi.org/10.1103/PhysRevD.94.052010}{\emph{Phys. Rev. D} {\bfseries 94} (2016) 052010} [\href{https://arxiv.org/abs/1606.07538}{{\ttfamily 1606.07538}}].

\bibitem{Kneale:2022sht}
L.~Kneale, M.~Smy and M.~Malek, \emph{{Coincidence-based reconstruction for reactor antineutrino detection in gadolinium-doped Cherenkov detectors}}, \href{https://doi.org/10.1016/j.nima.2023.168375}{\emph{Nucl. Instrum. Meth. A} {\bfseries 1053} (2023) 168375} [\href{https://arxiv.org/abs/2210.10576}{{\ttfamily 2210.10576}}].

\bibitem{Hyper-Kamiokande:2018ofw}
{\scshape Hyper-Kamiokande} collaboration, \emph{{Hyper-Kamiokande Design Report}},  \href{https://arxiv.org/abs/1805.04163}{{\ttfamily 1805.04163}}.

\bibitem{WATCHMAN:2015lcq}
{\scshape WATCHMAN} collaboration, \emph{{The Physics and Nuclear Nonproliferation Goals of WATCHMAN: A WAter CHerenkov Monitor for ANtineutrinos}},  \href{https://arxiv.org/abs/1502.01132}{{\ttfamily 1502.01132}}.

\bibitem{Cokinos:1977zz}
D.~Cokinos and E.~Melkonian, \emph{{Measurement of the 2200 m/sec neutron-proton capture cross section}}, \href{https://doi.org/10.1103/PhysRevC.15.1636}{\emph{Phys. Rev. C} {\bfseries 15} (1977) 1636}.

\bibitem{Strumia:2003zx}
A.~Strumia and F.~Vissani, \emph{{Precise quasielastic neutrino/nucleon cross-section}}, \href{https://doi.org/10.1016/S0370-2693(03)00616-6}{\emph{Phys. Lett. B} {\bfseries 564} (2003) 42} [\href{https://arxiv.org/abs/astro-ph/0302055}{{\ttfamily astro-ph/0302055}}].

\bibitem{Tomalak:2025jtn}
O.~Tomalak, \emph{{Theory of inverse beta decay for reactor antineutrinos}},  \href{https://arxiv.org/abs/2512.07956}{{\ttfamily 2512.07956}}.

\bibitem{Bodek:2007ym}
A.~Bodek, S.~Avvakumov, R.~Bradford and H.~S. Budd, \emph{{Vector and Axial Nucleon Form Factors:A Duality Constrained Parameterization}}, \href{https://doi.org/10.1140/epjc/s10052-007-0491-4}{\emph{Eur. Phys. J. C} {\bfseries 53} (2008) 349} [\href{https://arxiv.org/abs/0708.1946}{{\ttfamily 0708.1946}}].

\end{thebibliography}\endgroup

\end{document}